\documentclass[journal,twoside,web]{IEEEtran}
\usepackage{cite}
\usepackage{xr}
\usepackage{xcolor}
\usepackage{amssymb}
\usepackage{graphicx}
\usepackage{subfigure}
\usepackage{amsmath}
\usepackage{amsfonts}
\usepackage{multirow}
\usepackage{enumerate}
\usepackage{mathtools}
\usepackage{algorithm}
\usepackage[hidelinks]{hyperref}
\usepackage{textcomp}
\usepackage[noend]{algpseudocode}

\newtheorem{proposition}{Proposition}

\begin{document}
\title{SiliCoN: Simultaneous Nuclei Segmentation and Color Normalization of Histological Images}
\author{Suman Mahapatra and Pradipta Maji 
	\thanks{
		The authors are with Biomedical Imaging and Bioinformatics Lab,
		Machine Intelligence Unit, Indian Statistical Institute,
		Kolkata, India. E-mail: \{suman.maha.cs@gmail.com, pmaji@isical.ac.in\} 
		(Corresponding author: Pradipta Maji).}
}

\maketitle

\begin{abstract}
Segmentation of nuclei regions from histological images is an important task for automated computer-aided analysis of histological images, particularly in the presence of impermissible color variation in the color appearance of stained tissue images. 
While color normalization enables better nuclei segmentation, accurate segmentation of nuclei structures makes color normalization rather trivial. 
In this respect, the paper proposes a novel deep generative model for simultaneously segmenting nuclei structures and normalizing color appearance of stained histological images.
This model judiciously integrates the merits of truncated normal distribution and spatial attention. The model assumes that the latent color appearance information, corresponding to a particular histological image, is independent of respective nuclei segmentation map as well as embedding map information. 
The disentangled representation makes the model
generalizable and adaptable as the modification or loss in color appearance information cannot be able to affect the nuclei segmentation map as well as embedding information.
Also, for dealing with the stain overlap of associated histochemical reagents, the prior for latent color appearance code is assumed to be a mixture of truncated normal distributions. The proposed model incorporates the concept of spatial attention for segmentation of nuclei regions from histological images. The performance of the proposed approach, along with a comparative analysis with related state-of-the-art algorithms, has been demonstrated on publicly available standard histological image data sets.
\end{abstract}


\section{Introduction}
\label{intro}
\PARstart{O}{ne} of the foremost tasks in digital pathology is segmentation of nuclei regions from histological images as nuclei structures provide significant morphological information, which aids in therapeutic diagnosis of underlying diseases. However, as nuclei structures can exhibit different color, morphologies, texture or can be occluded partially by different biological components and other nuclei structures, nuclei segmentation becomes a very critical task in histological image analysis. In practical scenario, nuclei segmentation becomes difficult as histological images exhibit impermissible variation in the color appearance of stained pathology images due to the involvement of different factors, such as, inconsistency in staining routine, storage condition, specimen width, and so on.

\par Color normalization is a procedure which tries to reduce the variation in color appearance among similar biological components within and between the images in a histological image set while retaining the histological and structural information present within the images. In \cite{Reinhard2001}, each RGB histological image is transformed into decorrelated L$\alpha\beta$ space \cite{Ruderman1998}, and based on the channel statistics of template image, channels of source image are standardized. In \cite{Macenko2009}, a plain fitting (PF) approach based on singular value decomposition (SVD) was utilized to compute stain representative vectors corresponding to each individual image. One of the main disadvantages of PF based methods is that the associated parameters cannot be computed adaptively. In \cite{Li2015}, corresponding to each image, a saturation-weighted (SW) hue histogram is computed, and then $k$-means clustering is applied on the computed SW hue histogram to estimate stain-specific vectors. The blind stain separation method in \cite{Vahadane2016} modeled each image pixel as a sparse mixture model of the involved stains. To impose sparsity and ensure that the factor matrices are non-negative, sparse non-negative matrix factorization (SNMF) has been used. However, the main drawback with non-negative matrix factorization (NMF) is that it suffers from unstable convergence problem.

\par In recent years, deep learning  has been regarded as a powerful
tool in the analysis of medical images. A task-specific discriminative 
model was proposed in \cite{Bentaieb2018} to extract stain-specific information. 
However, the performance of the model depends 
on the labor-intensive task that needs extra manual labeling efforts. 
A deep generative model based on generative adversarial network (GAN) was introduced 
in \cite{Zanjani2018} to extract the color appearance information of associated histochemical stains. In StainGAN method \cite{Shaban2019}, being motivated by the unpaired CycleGAN model \cite{Zhu2017}, a cycle-consistency loss term is incorporated in the 
respective objective function. 
The major drawbacks of the aforementioned
methods are that they fail to address the uncertainty attributed 
by the overlapping nature of histochemical reagents, and are unable to
extract color appearance as well as stain
bound information from each individual histological image. 
A circular clustering algorithm, based on the paradigm of expectation-maximization (EM), 
was proposed in \cite{Li2017} to compute the color concentration
matrix. In \cite{Maji2020}, the von Mises distribution based rough-fuzzy circular 
clustering ($\text{RFCC}_{\text{vM}}$) algorithm was proposed to address the color 
normalization problem. 
However, the approaches proposed in \cite{Li2017} and \cite{Maji2020} do not consider the 
correlation between stain bound and color appearance information, extracted from each 
histological image. 
Furthermore, these methods depend on NMF, which produces inconsistent color appearance 
matrix due to the unstable convergence problem associated with NMF.

\par On the other hand, nuclei segmentation focuses on the 
labeling of each individual nucleus, and distinguishing 
each nucleus from the background, other nuclei structures and biological components in a histological image. 
In \cite{Ronneberger2015}, the U-Net model has been proposed where, in the traditional encoder-decoder framework, the notion of skip connections is incorporated to extract low-level 
semantic information from images. 
The U-Net++ model proposed in \cite{Zhou2018} achieves feature propagation 
via dense interconnected skip connections.
However, the major limitation of the U-Net based models is their inability to 
distinguish clustered nuclei structures and high sensitivity to pre-specified 
associated parameters. 
In \cite{He2017}, for detecting objects from both natural and biomedical images, 
a two-stage object detection method, named Mask-R-CNN, was 
proposed. Corresponding to each nucleus, bounding boxes with different probability values are predicted and the bounding box 
with maximum probability value is selected to segment the nuclei region. 
In \cite{Graham2019}, a simultaneous segmentation and classification model, named HoVer-Net, 
has been proposed for simultaneously segmenting and classifying the nuclei regions from histological images. The concepts of horizontal and 
vertical distance maps were utilized in the HoVer-Net model for separating clustered or occluded nuclei structures. Most of these aforementioned supervised methods 
require a huge amount of annotated images, which are often practically 
infeasible to get as labeling each and every nucleus within a histological image needs lots of 
time, effort and moreover, expert knowledge. In Stardist model \cite{Weigert2022}, 
a nuclei instance segmentation and classification approach was proposed. The Stardist model fails miserably to segment heterogeneous nuclei shapes as Stardist works on the assumption 
that the objects to be segmented are star-convex in shape, which is an invalid assumption 
regarding nuclei structures corresponding to different organs. 
Recently, a type of vision transformer, named Swin transformer, has emerged 
as a new tool for image segmentation. 
In \cite{Qian2022}, a Swin vision transformer based multiple instance learning model, termed as Swin-MIL, has been proposed for the prediction of nuclei segmentation masks from 
histological images. 
However, the vision transformer based models are computationally very expensive as they contain a large number of parameters and also need a substantial number of samples 
for training the models. 
In a recent approach, called BoNuS \cite{Lin2024}, a binary mining loss has 
been introduced for simultaneously learning nuclei interior and extracting boundary 
details to segment nuclei regions from histological images. 

\par One of the main highlights of \cite{Mahapatra2023} was to show that stain color normalization acts as an important preprocessing task, that eventually enables better segmentation of nuclei regions from histological images. Accurate nuclei segmentation, on the other hand, makes normalizing color appearance of stained tissue rather trivial. Hence, color normalization and nuclei segmentation can be regarded as two intertwined procedures. 
So, the main intuition is that these procedures should be integrated together for simultaneously yielding better color normalization and nuclei segmentation as the procedures can take advantage from each other during training. Although, in some recent 
works \cite{Prusty2023}, \cite{Martos2023}, \cite{Mahbod2024}, the problem of nuclei segmentation followed by color normalization, is addressed, the problem of simultaneously segmenting nuclei regions and normalizing color appearance is still not explored in the literature. 

\par In this context, the paper proposes a novel model, named \underline{Si}multaneous Nuc\underline{l}ei Segmentat\underline{i}on and \underline{Co}lor \underline{N}ormalization (SiliCoN). It judiciously integrates the merits of truncated normal distribution and spatial attention. The proposed SiliCoN model works based on the assumption that the latent color 
appearance code, corresponding to a particular histological image, is independent of respective nuclei segmentation map as well as embedding map. 
While the nuclei segmentation map captures information regarding the nuclei regions, the 
embedding map extracts details regarding cytoplasm and other cellular components. 
The disentangled representation ensures that the modification or loss in latent color 
appearance information does not impact the nuclei segmentation map as well as embedding map. 
Since the outer tails of a mixture of probability distributions are susceptible to outliers 
and also do not have sufficient 
contribution in handling stain overlap, the SiliCoN model assumes the prior for latent color appearance code to be a mixture of truncated normal distributions to address the overlapping nature of associated stains. The concept of spatial attention is incorporated in the proposed framework to extract segmentation maps corresponding to nuclei regions within histological images. The efficacy of the SiliCoN model in both stain color normalization and nuclei segmentation is demonstrated using benchmark Hematoxylin and Eosin (H\&E) stained histological image sets. 
Some of the results presented in this paper were originally reported in the doctoral dissertation of the first author \cite{Mahapatra2024}.

\section{SiliCoN: Proposed Method}

\label{proposed}
A deep generative model, named SiliCoN, is introduced in this section for simultaneously segmenting nuclei regions and normalizing color appearance of histological images.

\subsection{Problem Definition}
\label{SiliCoN_problem}
Consider an image data set $\mathbb{X}$, that contains $n$ number of 
histological images $\{x_{i}: i = 1, 2, \cdots, n\}$, the aim is to develop a model
that takes as input a non-normalized histological image 
$x_{i} \in \mathbb{X} \subset \mathbb{R}^{H\times W\times 3}$ and respective nuclei 
segmentation map $y_{i}$ is obtained as output. These output segmentation 
maps $\{y_{i}\}$ constitute the segmentation map domain $\mathbb{Y}$, that is, 
$y_{i} \in \mathbb{Y} \subset \mathbb{R}^{H\times W}$. The segmentation map $y_{i}$ 
is fed to the model as input and corresponding to the input histological image $x_{i}$, 
the stain color normalized image $\hat{x}_{i}$ is obtained as output. It can be observed that 
there is an information-imbalance between two image spaces: histological image space $\mathbb{X}$, which is an information-rich domain as each histological image contains 
information regarding cell nuclei and other cellular details, and information-poor 
segmentation map domain $\mathbb{Y}$, where each segmentation map image contains 
information regarding cell nuclei only. Color variation can also be observed among any two 
images $x_{i}$ and $x_{j}$, $i \neq j$, chosen randomly from image space $\mathbb{X}$, due to several factors responsible for variation in color appearance among 
histological images. Hence, developing a deep 
generative framework, which is capable of generating stain color 
normalized image corresponding to input non-normalized histological tissue image, and nuclei segmentation map, corresponding to the generated color normalized image simultaneously by addressing the information-imbalance between two 
asymmetric image spaces, is the ultimate goal of this study.

\begin{figure*}[th]
	\centerline{\includegraphics[width=0.875\textwidth]{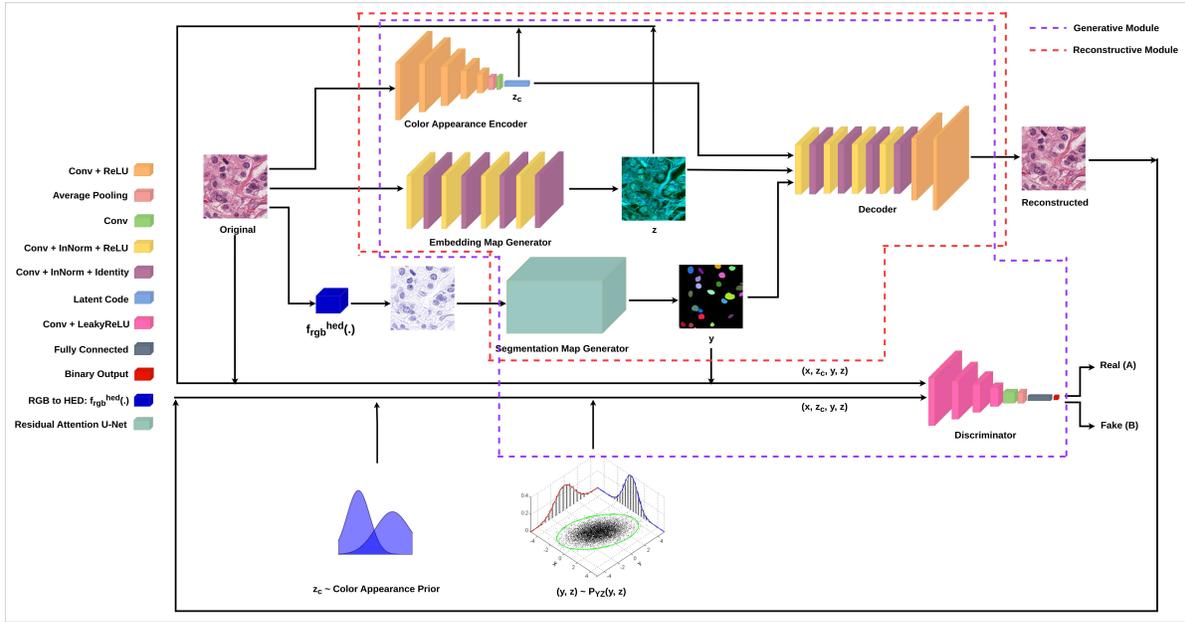}}
	\caption{Block diagram of the proposed SiliCoN model to simultaneously segment nuclei structures from and normalize color appearance of histological images. In this diagram, `Conv' and `InNorm' represent convolutional layer and instance normalization, respectively, and 'Identity' denotes identity function, i.e., $f(x) = x$.
	}
	\label{SiliCoN_block_diagram}
	\vspace{-0.32cm}
\end{figure*}

\subsection{Model Structure}
\label{SiliCoN_model}

The block diagram of several components constituting the proposed deep generative model, termed as SiliCoN, is presented in Fig. \ref{SiliCoN_block_diagram}. It is clear from Fig. \ref{SiliCoN_block_diagram} that the SiliCoN model is composed of five 
deep neural networks: $\mathbb{E}_{c}$, a color appearance encoder, for encoding the 
information regarding color appearance corresponding to a histological image, a segmentation map 
generator $\mathbb{F}_{\phi}$, that generates segmentation map corresponding 
to nuclei regions of each histological image, an embedding map generator $\mathbb{E}_{\omega}$ for encoding the additional information during the generation of information-poor segmentation map domain from information-rich histological image space, a decoder $\mathbb{G}$ and a discriminator $\mathbb{D}$. Let, the associated network parameters
corresponding to deep networks $\mathbb{E}_{c}$, $\mathbb{F}_{\phi}$, 
$\mathbb{E}_{\omega}$, $\mathbb{G}$ and $\mathbb{D}$ be represented by 
$\Theta_{\mathbb{E}_{c}}$, $\Theta_{\mathbb{F}_{\phi}}$, $\Theta_{\mathbb{E}_{\omega}}$, $\Theta_{\mathbb{G}}$ and $\Theta_{\mathbb{D}}$, respectively.
Each individual training histological image sample $x \in \mathbb{X}$ 
is simultaneously fed into the encoder and map generators: color appearance encoder $\mathbb{E}_{c}(x;\Theta_{\mathbb{E}_{c}})$, segmentation map generator $\mathbb{F}_{\phi}(x;\Theta_{\mathbb{F}_{\phi}})$ and embedding map generator $\mathbb{E}_{\omega}(x;\Theta_{\mathbb{E}_{\omega}})$ to eventually obtain corresponding 
latent color appearance code $z_{c}$, nuclei segmentation map 
$y$ and embedding map $z_{\omega}$, respectively. The nuclei segmentation map $y$ and embedding information $z_{\omega}$, along with latent color appearance code $z_{c}$, are then fed as inputs 
into the decoder $\mathbb{G}(z_{c}, y, z_{\omega}; \Theta_{\mathbb{G}})$ and $\mathbb{G}(z_{c}, y, z_{\omega})$ generates stain color normalized histological image $\hat{x}$ corresponding to the non-normalized input image $x$. 
Inputs in quadruplet form $(x, z_{c}, y, z_{\omega})$ are fed into the discriminator $\mathbb{D}(x, z_{c}, y, z_{\omega}; \Theta_{\mathbb{D}})$ and $\mathbb{D}$ distinguishes real data encoding, where
input image $x$ is sampled from original data distribution, $z_{c}$, $y$ and $z_{\omega}$ correspond to encoded color appearance information through $\mathbb{E}_{c}$, generated segmentation map using 
$\mathbb{F}_{\phi}$, and embedding information  obtained through 
$\mathbb{E}_{\omega}$, respectively, 
from fake data encoding, where the reconstructed image patch $\hat{x}$ is fed as $x$ and
$z_{c}$, $y$ and $z_{\omega}$ are sampled from corresponding prior distributions. 
These deep networks $\mathbb{E}_{c}$, $\mathbb{F}_{\phi}$, $\mathbb{E}_{\omega}$, 
$\mathbb{G}$ and $\mathbb{D}$ must be differentiable non-linear functions in order to ensure the back-propagation of the error values during the training phase. 
The nuclei regions in each H\&E-stained histological image are 
highlighted by H stain. Based on this information, the H-channel of a 
histological image $x \in \mathbb{X}$ is extracted via the operation 
$f_{rgb}^{hed}(x)[:, :, 0]$, where $f_{rgb}^{hed}(.)$ represents a function that takes 
as input H\&E-stained RGB histological image and outputs respective image in 
Hematoxylin-Eosin-DAB (HED) color space. A spatial attention network $\mathbb{F}_{\phi}$ takes the extracted H-channel image as input to extract nuclei regions corresponding to the H-channel image. In this study, the proposed deep generative framework incorporates spatial attention through existing residual attention U-Net model proposed in \cite{Jin2020}. 

\subsection{Model Objective}
The proposed SiliCoN model works with the objective of extracting latent color appearance
code, which is independent of nuclei segmentation map as well as generated 
embedding map, and learns to generate fake data encoding to be in close proximity to the 
real data encoding by optimizing the following minimization objective:
\begin{equation}
\label{SiliCoN_ultimate_objective}
J_\mathrm{Total} = \lambda_{Adv} \times J_\mathrm{Adv} +
\lambda_{Rec} \times J_\mathrm{Rec},
\end{equation}
where the objective terms $J_\mathrm{Adv}$
and $J_\mathrm{Rec}$, to be minimized, correspond to the generation module and the
reconstruction module, respectively, that are demonstrated next. The relative importance of the aforementioned terms $J_\mathrm{Adv}$ and $J_\mathrm{Rec}$ are represented by $\lambda_{Adv}$
and $\lambda_{Rec}$, respectively. Here, in this study, $J_{\rm Total}$ is computed by the 
convex combination 
of terms $J_{\rm Adv}$ and $J_{\rm Rec}$, that is, $\lambda_{Adv} + \lambda_{Rec} = 1$.  

\subsubsection{Generation Module}
As marked with the violet dashed line in Fig. \ref{SiliCoN_block_diagram}, all five deep neural networks: $\mathbb{E}_{c}$,
$\mathbb{F}_{\phi}$, $\mathbb{E}_{\omega}$, $\mathbb{G}$ and $\mathbb{D}$ constitute the generation module. 
This module is developed based on the assumption that the 
latent color appearance code is independent of the generated nuclei segmentation map 
as well as the embedding map. Accordingly, the generation module attributes the objective functions, that are formulated using the following expressions:
\begin{equation}
\label{SiliCoN_objective_disc}
J_{\mathrm{G}}(\mathbb{D}) = \min_{\mathbb{D}} J_{1}(\mathbb{E}_{c}, \mathbb{F}_{\phi}, \mathbb{E}_{\omega}, \mathbb{G}, \mathbb{D}), ~~~\text{where }
\end{equation}
\begin{equation*}
J_{1}(\mathbb{E}_{c}, \mathbb{F}_{\phi}, \mathbb{E}_{\omega}, \mathbb{G}, \mathbb{D}) = $$ $$ 
\underbrace{\begin{aligned} E_{x \sim P_{\mathbb{X}}(x)} E_{z_{c} \sim P_{\mathbb{E}_{c}}(z_{c} \mid x)}
	E_{y\sim P_{\mathbb{F}_{\phi}}(y \mid x)} E_{z_{\omega}\sim P_{\mathbb{E}_{\omega}}(z_{\omega} \mid x)} \\
	(\mathbb{D}[x, z_{c}, y, z_{\omega}] - A)^2 \end{aligned}}_\text{$R$} $$ $$
+\underbrace{\begin{aligned} E_{z_{c} \sim P_{\mathbb{Z}_{c}}(z_{c})} E_{y \sim P_{\mathbb{Y}}(y)} E_{z_{\omega} \sim P_{\mathbb{Z_{\omega}}}(z_{\omega})}
	E_{x \sim P_{\mathbb{G}}(x \mid z_{c}, y, z_{\omega})} \\ 
	(\mathbb{D}[x, z_{c}, y, z_{\omega}] - B)^2\end{aligned}}_\text{$F$}
\end{equation*}
Here, $A$ and $B$ are the labels, that the discriminator $\mathbb{D}$ assigns, to designate
the real data encoding and the fake data encoding, respectively. Also, the parts associated with real and fake 
data encoding are designated by the indicators $R$ and $F$, respectively, and
\begin{equation}
\label{SiliCoN_objective_gen}
J_{\mathrm{G}}(\mathbb{G}) = \min_{\mathbb{G}} J_{2}(\mathbb{E}_{c},  \mathbb{F}_{\phi}, \mathbb{E}_{\omega}, \mathbb{G}, \mathbb{D}), ~~~\text{where } $$ $$
J_{2}(\mathbb{E}_{c},  \mathbb{F}_{\phi}, \mathbb{E}_{\omega}, \mathbb{G}, \mathbb{D}) = $$ $$
E_{z_{c} \sim P_{\mathbb{Z}_{c}}(z_{c})} E_{y \sim P_{\mathbb{Y}}(y)} E_{z_{\omega} \sim P_{\mathbb{Z_{\omega}}}(z_{\omega})}
E_{x \sim P_{\mathbb{G}}(x \mid z_{c}, y, z_{\omega})} $$ $$
(\mathbb{D}[x, z_{c}, y, z_{\omega}] - C)^2
\end{equation}
Here, as desired by the decoder $\mathbb{G}$, discriminator $\mathbb{D}$ assigns the label $C$ to indicate the fake data encoding. 

\par In real data encoding part $R$ of (\ref{SiliCoN_objective_disc}), $P_{\mathbb{X}}(x)$ denotes the real data
distribution, and given the sample histological image patch $x$, $P_{\mathbb{E}_{c}}(z_{c} \mid x)$, $P_{\mathbb{F}_{\phi}}(y \mid x)$ and $P_{\mathbb{E}_{\omega}}(z_{\omega} \mid x)$ represent the
conditional distributions corresponding to networks $\mathbb{E}_{c}$, $\mathbb{F}_{\phi}$ and $\mathbb{E}_{\omega}$, respectively. In case of fake data encoding part $F$ of (\ref{SiliCoN_objective_disc}), the prior distributions, from which latent color appearance code $z_{c}$, nuclei segmentation map $y$ and embedding map
$z_{\omega}$ are sampled, are represented by $P_{\mathbb{Z}_{c}}(z_{c})$, $P_{\mathbb{Y}}(y)$ and $P_{\mathbb{Z_{\omega}}}(z_{\omega})$, respectively, and here, given the latent color appearance 
code $z_{c}$, and the generated maps $y$ and $z_{\omega}$, the conditional
distribution corresponding to reconstructed image patch  $\hat{x}$ is denoted by $P_{\mathbb{G}}(x\mid z_{c}, y, z_{\omega})$. So, by simplifying (\ref{SiliCoN_objective_disc}), the following can be written:
\begin{equation*}
	J_{1}(\mathbb{E}_{c}, \mathbb{F}_{\phi}, \mathbb{E}_{\omega}, \mathbb{G}, \mathbb{D}) =
	\int\limits_{x} P_{\mathbb{X}}(x) \int\limits_{z_{c}} P_{\mathbb{E}_{c}}(z_{c} \mid x)
	\int\limits_{y} P_{\mathbb{F}_{\phi}}(y \mid x) $$ $$
	\int\limits_{z_{\omega}} P_{\mathbb{E}_{\omega}}(z_{\omega} \mid x) 
	\times (\mathbb{D}[x, z_{c}, y, z_{\omega}] - A)^2 dz_{\omega} dy dz_{c} dx $$ $$
	+ \int\limits_{z_{c}} P_{\mathbb{Z}_{c}}(z_{c}) \int\limits_{y} P_{\mathbb{Y}}(y) \int\limits_{z_{\omega}} P_{\mathbb{Z_{\omega}}}(z_{\omega})
	\int\limits_{x} P_{\mathbb{G}}(x \mid z_{c}, y, z_{\omega}) $$ $$
	(\mathbb{D}[x, z_{c}, y, z_{\omega}] - B)^2
	dx dz_{\omega} dy dz_{c} $$ $$
	= \int\limits_{\{x, z_{c}, y, z_{\omega}\}} [P_{\mathbb{X}}(x) P_{\mathbb{E}_{c}}(z_{c} \mid x)
	P_{\mathbb{F}_{\phi}}(y \mid x)
	P_{\mathbb{E}_{\omega}}(z_{\omega} \mid x) $$ $$ 
	(\mathbb{D}[x, z_{c}, y, z_{\omega}] - A)^2 
	+ P_{\mathbb{Z}_{c}}(z_{c}) P_{\mathbb{Y}}(y) P_{\mathbb{Z_{\omega}}}(z_{\omega})
	P_{\mathbb{G}}(x \mid z_{c}, y, z_{\omega}) $$ $$
	(\mathbb{D}[x, z_{c}, y, z_{\omega}] - B)^2] dx dz_{c} dy dz_{\omega} $$ $$
	= \int\limits_{\{x, z_{c}, y, z_{\omega}\}} [P_{\mathbb{X}}(x) P_{\mathbb{E}_{c} \mathbb{F}_{\phi} \mathbb{E}_{\omega}}(z_{c}, y, z_{\omega} \mid x)
	(\mathbb{D}[x, z_{c}, y, z_{\omega}] - A)^2 + $$ $$
	P_{\mathbb{Z}_{c} \mathbb{Y} \mathbb{Z_{\omega}}}(z_{c}, y, z_{\omega}) P_{\mathbb{G}}(x \mid z_{c}, y, z_{\omega})
	(\mathbb{D}[x, z_{c}, y, z_{\omega}] - A)^2] dx dz_{c} dy dz_{\omega}
\end{equation*}
as per the assumption, latent code $z_{c}$ is independent of both segmentation map $y$ and 
embedding information $z_{\omega}$. Hence, 
\begin{align}
\label{SiliCoN_gen_obj_simple}
&J_{1}(\cdot)
=\int\limits_{\{x, z_{c}, y, z_{\omega}\}} [P_{\mathbb{E}_{c} \mathbb{F}_{\phi} \mathbb{E}_{\omega} \mathbb{X}}
(x, z_{c}, y, z_{\omega}) (\mathbb{D}[x, z_{c}, y, z_{\omega}] - A)^2 \nonumber\\
& + P_{\mathbb{G} \mathbb{Z}_{c} \mathbb{Y} \mathbb{Z_{\omega}}}(x, z_{c}, y, z_{\omega}) (\mathbb{D}[x, z_{c}, y, z_{\omega}] - B)^2 ] dx dz_{c} dy dz_{\omega}.
\end{align}
\par Now, given any combination of encoder, map generators and decoder ($\mathbb{E}_{c}$, $\mathbb{F}_{\phi}$, $\mathbb{E}_{\omega}$, $\mathbb{G}$), the first task is to compute the optimal discriminator 
$\mathbb{D}^{*}$.

\begin{proposition}
	The optimal discriminator $\mathbb{D}^{*}$, corresponding to a fixed combination of color appearance encoder $\mathbb{E}_{c}$, segmentation 
	map generator $\mathbb{F}_{\phi}$, embedding map generator $\mathbb{E}_{\omega}$ and decoder $\mathbb{G}$, is expressed as follows:
	\begin{equation}
		\mathbb{D}^{*}[x, z_{c}, y, z_{\omega}] = $$ $$
		\frac{A. P_{\mathbb{E}_{c}\mathbb{F}_{\phi} \mathbb{E}_{\omega}\mathbb{X}}(x, z_{c}, y, z_{\omega}) +
			B. P_{\mathbb{G}\mathbb{Z}_{c} \mathbb{Y} \mathbb{Z_{\omega}}}(x, z_{c}, y, z_{\omega})}
		{P_{\mathbb{E}_{c} \mathbb{F}_{\phi} \mathbb{E}_{\omega}\mathbb{X}}(x, z_{c}, y, z_{\omega}) +
			P_{\mathbb{G}\mathbb{Z}_{c}\mathbb{Y} \mathbb{Z_{\omega}}}(x, z_{c}, y, z_{\omega})}
	\end{equation}
\end{proposition}

\par \textit{Proof.} 
For learning the discriminator $\mathbb{D}$, corresponding to a fixed combination of $\mathbb{E}_{c}$, $\mathbb{F}_{\phi}$, 
$\mathbb{E}_{\omega}$ and $\mathbb{G}$, the objective term $J_{1}(\mathbb{E}_{c}, \mathbb{E}_{s}, \mathbb{G}, \mathbb{D})$ needs to be minimized by differentiating $J_{1}$ partially with respect to $\mathbb{D}$ as follows:
\begin{align}
&\quad \quad \quad \quad \frac{\partial J_{1}(\mathbb{E}_{c}, \mathbb{F}_{\phi}, \mathbb{E}_{\omega}, 
\mathbb{G}, \mathbb{D})}
{\partial \mathbb{D}[x, z_{c}, y, z_{\omega}]} = \nonumber\\
&2 P_{\mathbb{E}_{c}\mathbb{F}_{\phi}\mathbb{E}_{\omega}\mathbb{X}}(x, z_{c}, y, z_{\omega})
(A- \mathbb{D}[x, z_{c}, y, z_{\omega}]) \nonumber\\
&\quad \quad + 2 P_{\mathbb{G}\mathbb{Z}_{c}\mathbb{Y}\mathbb{Z_{\omega}}}(x, z_{c}, y, z_{\omega})
(B - \mathbb{D}[x, z_{c}, y, z_{\omega}])
\label{SiliCoN_DerivateDisc}
\end{align}
\par Let, $\mathbb{D}^{*}[x, z_{c}, y, z_{\omega}]$ represent the optimal discriminator.
So,
\begin{equation*}
\frac{\partial J_{1}(\mathbb{E}_{c}, \mathbb{F}_{\phi}, \mathbb{E}_{\omega}, 
\mathbb{G}, \mathbb{D})}
{\partial \mathbb{D}[x, z_{c}, y, z_{\omega}]}
\bigg|_{\mathbb{D} = \mathbb{D}^{*}} = 0; $$ $$
~~~~~\Rightarrow 2 \{P_{\mathbb{E}_{c}\mathbb{F}_{\phi}\mathbb{E}_{\omega}\mathbb{X}}
(x, z_{c}, y, z_{\omega})
(A - \mathbb{D}^{*}[x, z_{c}, y, z_{\omega}]) +
$$ $$ ~~~~~~ P_{\mathbb{G}\mathbb{Z}_{c}\mathbb{Y}\mathbb{Z_{\omega}}}(x, z_{c}, y, z_{\omega})
(B - \mathbb{D}^{*}[x, z_{c}, y, z_{\omega}])\} = 0; 
\text{[from (\ref{SiliCoN_DerivateDisc})]} $$ $$
\Rightarrow \mathbb{D}^{*}[x, z_{c}, y, z_{\omega}] =
\end{equation*}
\begin{eqnarray}
	\frac{A. P_{\mathbb{E}_{c}\mathbb{F}_{\phi}\mathbb{E}_{\omega}\mathbb{X}}
			(x, z_{c}, y, z_{\omega}) +
		{B. P_{\mathbb{G}\mathbb{Z}_{c}\mathbb{Y}\mathbb{Z_{\omega}}}(x, z_{c}, y, z_{\omega})}}
	{P_{\mathbb{E}_{c}\mathbb{F}_{\phi}\mathbb{E}_{\omega}\mathbb{X}}(x, z_{c}, y, z_{\omega}) +
		P_{\mathbb{G}\mathbb{Z}_{c}\mathbb{Y}\mathbb{Z_{\omega}}}(x, z_{c}, y, z_{\omega})}
	\label{OptimalDisc}
\end{eqnarray}

\par Theoretically, if $(\mathbb{E}_{c}, \mathbb{F}_{\phi}, \mathbb{E}_{\omega}, \mathbb{G})^{*}$ represents optimal combination of encoder, map generators and decoder,
then the discriminator must not be capable of distinguishing real data encoding from
fake data encoding due to the fact that decoder $\mathbb{G}$, upon training, becomes able to mimic the intrinsic real joint distribution. So, the optimal discriminator $\mathbb{D}^{*}$, at equilibrium point, cannot be able to discriminate real data encoding from fake data encoding, and use same label $C$, that the decoder $\mathbb{G}$ tricks
discriminator $\mathbb{D}$ to believe, to designate both the encodings. Thus, the following can be written from (\ref{SiliCoN_gen_obj_simple}):

\begin{equation*}
	(\mathbb{E}_{c}, \mathbb{F}_{\phi}, \mathbb{E}_{\omega}, \mathbb{G})^{*}  = \arg\min_{\mathbb{E}_{c}, \mathbb{F}_{\phi}, \mathbb{E}_{\omega}, 
		\mathbb{G}} J_{2}
	(\mathbb{E}_{c}, \mathbb{F}_{\phi}, \mathbb{E}_{\omega}, \mathbb{G}, \mathbb{D}^{*}), \text{where }$$ $$
	J_{2}(\mathbb{E}_{c}, \mathbb{F}_{\phi}, \mathbb{E}_{\omega}, \mathbb{G}, \mathbb{D}^{*}) = 
	$$ $$
	E_{x \sim P_{\mathbb{X}}(x)} E_{z_{c} \sim P_{\mathbb{E}_{c}}(z_{c} \mid x)}
	E_{y \sim P_{\mathbb{F}_{\phi}}(y \mid x)} E_{z_{\omega} \sim P_{\mathbb{E}_{\omega}}(z_{\omega} \mid x)} $$ $$
	(\mathbb{D}^{*}[x, z_{c}, y, z_{\omega}] - C)^2 $$ $$
	+E_{z_{c} \sim P_{\mathbb{Z}_{c}}(z_{c})} E_{y \sim P_{\mathbb{Y}}(y)} E_{z_{\omega} \sim P_{\mathbb{Z_{\omega}}}(z_{\omega})}
	E_{x \sim P_{\mathbb{G}}(x \mid z_{c}, y, z_{\omega})} $$ $$
	(\mathbb{D}^{*}[x, z_{c}, y, z_{\omega}] - C)^2 $$ $$
	\text{[optimization does not get impacted by $\mathbb{G}$-independent terms]} $$ $$
	= \int\limits_{\{x, z_{c}, y, z_{\omega}\}} \{P_{\mathbb{E}_{c} \mathbb{F}_{\phi} \mathbb{E}_{\omega} \mathbb{X}}(x, z_{c}, y, z_{\omega})
	(\mathbb{D}^{*}[x, z_{c}, y, z_{\omega}] - C)^2 + $$ $$
	P_{\mathbb{G} \mathbb{Z}_{c} \mathbb{Y} \mathbb{Z_{\omega}}}(x, z_{c}, y, z_{\omega}) (\mathbb{D}^{*}[x, z_{c}, y, z_{\omega}] - C)^2\} dx dz_{c} dy dz_{\omega} $$ $$
	= \int\limits_{\{x, z_{c}, y, z_{\omega}\}} \{P_{\mathbb{E}_{c} \mathbb{F}_{\phi} \mathbb{E}_{\omega} \mathbb{X}}(x, z_{c}, y, z_{\omega}) + 
	P_{\mathbb{G} \mathbb{Z}_{c} \mathbb{Y} \mathbb{Z_{\omega}}}(x, z_{c}, y, z_{\omega})\} 
	$$ $$ \times
	\left[C - \frac{A. P_{\mathbb{E}_{c}\mathbb{F}_{\phi}\mathbb{E}_{\omega}\mathbb{X}}(x, z_{c}, y, z_{\omega}) +
		B. P_{\mathbb{G}\mathbb{Z}_{c}\mathbb{Y}\mathbb{Z_{\omega}}}(x, z_{c}, y, z_{\omega})}
	{P_{\mathbb{E}_{c}\mathbb{F}_{\phi}\mathbb{E}_{\omega}\mathbb{X}}(x, z_{c}, y, z_{\omega}) +
		P_{\mathbb{G}\mathbb{Z}_{c}\mathbb{Y}\mathbb{Z_{\omega}}}(x, z_{c}, y, z_{\omega})}\right]^{2} 
	$$ $$ dx dz_{c} dy dz_{\omega}.
\end{equation*}

\par So, the generative module shapes the overall adversarial objective, which can be computed as follows:
\begin{equation}
	\label{SiliCoN_adversarial_loss}
	J_\mathrm{Adv} = J_{\mathrm{G}}(\mathbb{D}) + J_{\mathrm{G}}(\mathbb{G}).
\end{equation}

\subsubsection{Reconstruction Module}
As highlighted by the red dashed line in Fig. \ref{SiliCoN_block_diagram}, four networks, $\mathbb{E}_{c}$,
$\mathbb{F}_{\phi}$, $\mathbb{E}_{\omega}$ and $\mathbb{G}$ constitute the reconstruction module. Again, the reconstruction module 
is also designed depending on the assumption that the color appearance code $z_{c}$ is independent of 
generated nuclei segmentation map $y$ as well as the embedding information $z_{\omega}$. Now, the decomposition 
of joint distribution $P_{\mathbb{G}\mathbb{Z}_{c}\mathbb{Y}\mathbb{Z_{\omega}}}$, associated with the reconstruction of image patches, is given as follows:
\begin{equation}
	P_{\mathbb{G}\mathbb{Z}_{c}\mathbb{Y}\mathbb{Z_{\omega}}}(x, z_{c}, y, z_{\omega}) = P_{\mathbb{G}}(x \mid z_{c}, y, z_{\omega})P_{\mathbb{Z}_{c}\mathbb{Y}\mathbb{Z_{\omega}}}(z_{c}, y, z_{\omega}).
\end{equation}
Now, the latent color appearance code $z_{c}$ is expected to be independent of 
both the generated maps $y$ and $z_{\omega}$. But, the segmentation map $y$ cannot be assumed to be independent of embedding map $z_{\omega}$ as both the generated maps capture complementary information from histological images. Hence,
\begin{equation}
	\label{SiliCoN_decomposition}
	P_{\mathbb{G}\mathbb{Z}_{c}\mathbb{Y}\mathbb{Z_{\omega}}}(x, z_{c}, y, z_{\omega}) = P_{\mathbb{G}}(x \mid z_{c}, y, z_{\omega})P_{\mathbb{Z}_{c}}(z_{c})
	P_{\mathbb{Y}\mathbb{Z_{\omega}}}(y, z_{\omega}).
\end{equation}
\par Here, given the histological image patch $x$, the joint conditional density of the latent code $z_{c}$, and generated maps $y$ and $z_{\omega}$ needs to be computed to solve the inference problem. So,

\begin{equation}
	\label{SiliCoN_inference}
	P(z_{c}, y, z_{\omega}\mid x) = \frac{P_{\mathbb{G}\mathbb{Z}_{c}\mathbb{Y}\mathbb{Z_{\omega}}}(x, z_{c}, y, z_{\omega})}{P(x)},
\end{equation}
where model evidence or marginal likelihood is represented by $P(x)$, which can be computed by marginalizing over $z_{c}$, $y$ and $z_{\omega}$ as follows:
\begin{equation}
	\label{SiliCoN_marginal}
	P(x) = \int\limits_{z_{c}} \int\limits_{y} \int\limits_{z_{\omega}} P_{\mathbb{G}\mathbb{Z}_{c}\mathbb{Y}\mathbb{Z_{\omega}}}(x, z_{c}, y, z_{\omega})
	dz_{c} dy dz_{\omega}.
\end{equation}


\par Now, as the integral in (\ref{SiliCoN_marginal}) needs integration over multi-dimensional variables: latent representation
$z_{c}$, and generated maps $y$ and $z_{\omega}$, the computation 
of model evidence $P(x)$ generally becomes intractable. So,
a surrogate posterior distribution $Q(z_{c}, y, z_{\omega})$, that has a closed form solution and is easy to work with, needs to be utilized so that the posterior $P(z_{c}, y, z_{\omega}\mid x)$ can be approximated  with the help of surrogate posterior $Q(z_{c}, y, z_{\omega})$ by minimizing the Kullback-Leibler (KL) divergence between distributions $Q(z_{c}, y, z_{\omega})$ and $P(z_{c}, y, z_{\omega}\mid x)$ as follows:

\begin{align}
\label{SiliCoN_var_approx}
&\quad \quad \quad \quad \quad D_{KL}[Q(z_{c}, y, z_{\omega}) \mid\mid P(z_{c}, y, z_{\omega}\mid x)] \nonumber\\
&= - \int\limits_{z_{c}} \int\limits_{y} \int\limits_{z_{\omega}} Q(z_{c}, y, z_{\omega}) \log
\left[\frac{P(z_{c}, y, z_{\omega}\mid x)}{Q(z_{c}, y, z_{\omega})}\right] dz_{c} dy dz_{\omega} \nonumber\\
&= - E_{Q(z_{c}, y, z_{\omega})}\left[\log \left(\frac{P_{\mathbb{G}\mathbb{Z}_{c}\mathbb{Y}\mathbb{Z_{\omega}}}(x, z_{c}, y, z_{\omega})}
{Q(z_{c}, y, z_{\omega})}\right)\right] + \log P(x).
\end{align}

\par Now, the lower bound of KL divergence between two probability distributions is $0$. So, the following can be deduced from (\ref{SiliCoN_var_approx}):
\begin{align}
\label{SiliCoN_ELBO_eqn}
&-E_{Q(z_{c}, y, z_{\omega})}\left[\log \left( \frac{P_{\mathbb{G}\mathbb{Z}_{c}\mathbb{Y}\mathbb{Z_{\omega}}}(x, z_{c}, y, z_{\omega})}
{Q(z_{c}, y, z_{\omega})}\right)\right] + \log P(x) \ge 0 \nonumber\\
&\Rightarrow \log P(x) \ge E_{Q(z_{c}, y, z_{\omega})}\left[\log \left(\frac{P_{\mathbb{G}\mathbb{Z}_{c}\mathbb{Y}\mathbb{Z_{\omega}}}
(x, z_{c}, y, z_{\omega})}{Q(z_{c}, y, z_{\omega})}\right)\right].
\end{align}
\par Here, $P(x)$ being the model evidence, the log evidence is represented by $\log P(x)$  and 
hence, the right hand side of (\ref{SiliCoN_ELBO_eqn}) denotes the evidence lower bound (ELBO). So, it can be stated 
that maximizing the
ELBO in (\ref{SiliCoN_ELBO_eqn}) is equivalent to minimizing the KL divergence in (\ref{SiliCoN_var_approx}).

\par Now, taking negative of (\ref{SiliCoN_ELBO_eqn}), the following upper bound can be obtained:
\begin{equation}
	\label{SiliCoN_negativelog}
	-\log P(x) \le -E_{Q(z_{c}, y, z_{\omega})}\left[\log\left(
	\frac{P_{\mathbb{G}\mathbb{Z}_{c}\mathbb{Y}\mathbb{Z_{\omega}}}(x, z_{c}, y, z_{\omega})}{Q(z_{c}, y, z_{\omega})}\right)\right].
\end{equation}

Let, the minimization term, represented by the right hand side of (\ref{SiliCoN_negativelog}), be denoted by $\mathbb{P}$. Now,
\begin{equation*}
	\mathbb{P} = -E_{Q(z_{c}, y, z_{\omega})}\left[\log\left(
	\frac{P_{\mathbb{G}\mathbb{Z}_{c}\mathbb{Y}\mathbb{Z_{\omega}}}(x, z_{c}, y, z_{\omega})}{Q(z_{c}, y, z_{\omega})}\right)\right].
\end{equation*}
Using (\ref{SiliCoN_decomposition}), $\mathbb{P}$ can be rewritten as follows:
\begin{equation*} 
	\mathbb{P} = -E_{Q(z_{c}, y, z_{\omega})}\left[\log
	\frac{P_{\mathbb{G}}(x \mid z_{c}, y, z_{\omega})P_{\mathbb{Z}_{c}}(z_{c})
		P_{\mathbb{YZ_{\omega}}}(y, z_{\omega})}{Q(z_{c}, y, z_{\omega})}\right] $$ $$
	= -E_{Q(z_{c}, y, z_{\omega})}[\log P_{\mathbb{G}}(x \mid z_{c}, y, z_{\omega})] $$ $$
	-E_{Q(z_{c}, y, z_{\omega})} \left[\log\frac{P_{\mathbb{Z}_{c}}(z_{c})}
	{Q(z_{c}, y, z_{\omega})}\right] $$ $$
	-E_{Q(z_{c}, y, z_{\omega})} \left[\log\frac{P_{\mathbb{YZ_{\omega}}}(y, z_{\omega})}
	{Q(z_{c}, y, z_{\omega})}\right]
	-E_{Q(z_{c}, y, z_{\omega})} [\log Q(z_{c}, y, z_{\omega})]
\end{equation*}
\begin{align}
\label{SiliCoN_recon_obj}
&\quad \quad \quad= -E_{Q(z_{c}, y, z_{\omega})}[\log P_{\mathbb{G}}(x \mid z_{c}, y, z_{\omega})] \nonumber\\
&+ D_{KL}[Q(z_{c}) \mid\mid P_{\mathbb{Z}_{c}}(z_{c})] 
+ D_{KL}[Q(y, z_{\omega}) \mid\mid P_{\mathbb{YZ_{\omega}}}(y, z_{\omega})] \nonumber\\
&\quad \quad \quad \quad \quad \quad - E_{Q(z_{c}, y, z_{\omega})} [\log Q(z_{c}, y, z_{\omega})]. 
\end{align}

Here, the term $E_{Q(z_{c}, y, z_{\omega})} [\log Q(z_{c}, y, z_{\omega})]$ in (\ref{SiliCoN_recon_obj}) denotes the entropy over surrogate distribution $Q(z_{c}, y, z_{\omega})$, which acts as a regularizer for 
$Q(z_{c}, y, z_{\omega})$.
As latent color appearance code $z_{c}$ is assumed to be independent of both 
the generated maps $y$ and $z_{\omega}$, from optimization 
perspective,
optimizing $D_{KL}[Q(z_{c}, y, z_{\omega}) \mid\mid P_{\mathbb{Z}_{c}}(z_{c})]$ is same as
optimizing $D_{KL}[Q(z_{c}) \mid\mid P_{\mathbb{Z}_{c}}(z_{c})]$.
Similarly, $D_{KL}[Q(z_{c}, y, z_{\omega}) \mid\mid P_{\mathbb{YZ}}(y, z_{\omega})]$ can be optimized by optimizing
$D_{KL}[Q(y, z_{\omega}) \mid\mid P_{\mathbb{YZ}}(y, z_{\omega})]$. Thus, from (\ref{SiliCoN_recon_obj}),
the reconstruction objective term $\mathbb{J}_\mathrm{rec}$ to be minimized can be derived as:
\begin{equation*}
	\mathbb{J}_\mathrm{Rec} = \underbrace{-E_{Q(z_{c}, y, z_{\omega})}
		[\log P_{\mathbb{G}}(x \mid z_{c}, y, z_{\omega})]}_\text{$L_{R}$} $$ $$
	-E_{Q(z_{c}, y, z_{\omega})} [\log Q(z_{c}, y, z_{\omega})]
\end{equation*}
\begin{equation}
	+ \underbrace{D_{KL}[Q(z_{c}) \mid\mid P_{\mathbb{Z}_{c}}(z_{c})]}_\text{$R_{1}$}
	+ \underbrace{D_{KL}[Q(y, z_{\omega}) \mid\mid P_{\mathbb{YZ}}(y, z_{\omega})]}_\text{$R_{2}$},
	\label{SiliCoN_reconstruction_loss}
\end{equation}
where $L_{R}$ denotes the reconstruction loss term, whereas terms $R_{1}$ and $R_{2}$
represent the regularization terms corresponding to latent color appearance code $z_{c}$, 
and joint density of nuclei segmentation map $y$ and embedding map $z_{\omega}$, respectively.


\par It is evident from (\ref{SiliCoN_reconstruction_loss}) that the minimization problem in (\ref{SiliCoN_reconstruction_loss}) can be solved only if the two distributions $P_{\mathbb{Z}_{c}}(z_{c})$ and $P_{\mathbb{YZ_{\omega}}}(y, z_{\omega})$ are known. So, both the
priors for latent code $z_{c}$,
and the joint density of generated maps $y$ and $z_{\omega}$, need 
to be assumed. The latent
color appearance code, extracted from each individual stained histological image patch, 
should capture information regarding all the histochemical reagents used in the staining 
routine. 
Hence, the color appearance code $z_{c}$ is assumed to be sampled from a mixture of truncated normal distributions. As, after stain color normalization, the segmentation map and the embedding map information, extracted from a particular histological image, 
must be retained, the prior for the joint density of 
segmentation map $y$ and embedding map $z_{\omega}$ is assumed to be a standard normal distribution. 
To ensure that the histological information is contained after reconstruction, a loss term $l_{SSIM}(\mathbb{G}, \mathbb{E}_{c}, \mathbb{F}_{\phi}, \mathbb{E}_{\omega})$, based on structural similarity measure, is incorporated along with the reconstruction loss term $\mathbb{J}_\mathrm{rec}$, which is defined 
as follows:
\begin{align}
& \quad \quad \quad \quad \quad \quad	l_{SSIM}(\mathbb{G}, \mathbb{E}_{c}, \mathbb{F}_{\phi}, \mathbb{E}_{\omega}) \nonumber \\ 
& \quad \quad	= 1 - SSIM[x, \mathbb{G}(\mathbb{E}_{c}(x), \mathbb{F}_{\phi}(x), \mathbb{E}_{\omega}(x))]
\end{align}

So, the overall reconstruction loss, which has to be minimized, can be framed as follows:
\begin{align}
	&\quad J_\mathrm{Rec} = \mathbb{J}_\mathrm{Rec} + l_{SSIM}(\mathbb{G}, \mathbb{E}_{c}, \mathbb{F}_{\phi}, \mathbb{E}_{\omega}) 
\label{SiliCoN_reconstruction_full}	
\end{align}

Here, SSIM denotes the structural similarity index measure \cite{Zhou2004}. The algorithm for simultaneous nuclei segmentation and color 
normalization is presented in Algorithm \ref{SiliCoN_map_algo}.

\begin{algorithm}[h]
	\textbf{Input:} Trained network parameters $\{\Theta_{\mathbb{E}_{c}}, \Theta_{\mathbb{F}_{\phi}}, \Theta_{\mathbb{E}_{\omega}}, \Theta_{\mathbb{G}}\}$
	associated with networks $\mathbb{E}_{c}$, $\mathbb{F}_{\phi}$, $\mathbb{E}_{\omega}$ and $\mathbb{G}$, template image $x^{T}$ and set of $N$ non-normalized source images $\{x_{n}^{S}\}_{n=1}^{N}$. \\
	\textbf{Output:} Color normalized source images $\{\tilde{x}_{n}^{S}\}_{n=1}^{N}$ and nuclei segmentation map $\{\tilde{y}_{n}^{S}\}_{n=1}^{N}$.
	\begin{algorithmic}[1]
		\State Corresponding to the template image $x^{T}$, generate latent color appearance code $z_{c}^{T}$, and nuclei segmentation map $y^{T}$ and embedding information $z_{\omega}^{T}$ by using $\mathbb{E}_{c}(x^{T}; \Theta_{\mathbb{E}_{c}})$, $\mathbb{F}_{\phi}(x^{T}; \Theta_{\mathbb{F}_{\phi}})$ and $\mathbb{E}_{\omega}(x^{T}; \Theta_{\mathbb{E}_{\omega}})$, respectively.
		\For{each image patch in source image set $\{x_{n}^{S}\}_{n=1}^{N}$}
		\begin{itemize}
			\item Generate latent color appearance code $z_{c_{n}}^{S}$, nuclei segmentation map $y_{n}^{S}$ and embedding information $z_{{\omega}_{n}}^{S}$ via $\mathbb{E}_{c}(x_{n}^{S}; \Theta_{\mathbb{E}_{c}})$, $\mathbb{F}_{\phi}(x_{n}^{S}; \Theta_{\mathbb{F}_{\phi}})$ and $\mathbb{E}_{\omega}(x_{n}^{S}; \Theta_{\mathbb{E}_{\omega}})$, respectively.
			\item Feed the latent color appearance code $z_{c}^{T}$ corresponding to the template image $x^{T}$, the source image nuclei segmentation map $y_{n}^{S}$ and the embedding information $z_{{\omega}_{n}}^{S}$ to the decoder $\mathbb{G}$ and by using
			$\mathbb{G}(z_{c}^{T}, y_{n}^{S}, z_{{\omega}_{n}}^{S}; \Theta_{\mathbb{G}})$ generate normalized source image
			$\tilde{x}_{n}^{S}$.
			\item Feed the normalized source image $\tilde{x}_{n}^{S}$, generated in the previous step to the segmentation map generator $\mathbb{F}_{\phi}$ and by using $\mathbb{F}_{\phi}(\tilde{x}_{n}^{S}; \Theta_{\mathbb{F}_{\phi}})$ the model generates final nuclei segmentation map $\tilde{y}_{n}^{S}$.  
		\end{itemize}
		\EndFor
		\State Stop.
	\end{algorithmic}
	\caption{Algorithm for simultaneously segmenting nuclei structures and normalizing color appearance of histological images.}
	\label{SiliCoN_map_algo}
\end{algorithm}


\section{Performance Analysis}
\label{performance}
\label{SiliCoN_results}
The effectiveness of the proposed simultaneous nuclei segmentation and color normalization model is presented in this section. The performance of the proposed SiliCoN model is compared with that of
\begin{itemize}
	\item several existing approaches for stain vector estimation: 
	plane fitting (PF) \cite{Macenko2009}, HTN \cite{Li2015}, enhanced PF (EPF) \cite{McCann2014},
	structure-preserving color normalization (SPCN) \cite{Vahadane2016},
	expectation-maximization (EM) algorithm \cite{Li2017}, and rough-fuzzy
	circular clustering method based on von Mises distribution $\text{RFCC}_\text{vM}$ \cite{Maji2020}; 
	\item several existing color normalization methods: color transfer
	(ColTrans) \cite{Reinhard2001}, stain color description (SCD) \cite{Khan2014},
	SN-GAN model \cite{Zanjani2018}, StainGAN \cite{Shaban2019}, 
	AST model \cite{Bentaieb2018}, along with the methods PF
	\cite{Macenko2009}, HTN \cite{Li2015}, EPF \cite{McCann2014}, SPCN 
	\cite{Vahadane2016}, $\text{RFCC}_\text{vM}$ \cite{Maji2020} and 
	TredMiL \cite{Mahapatra2023}; and 
	\item several state-of-the-art deep models: 
	U-Net \cite{Ronneberger2015}, Mask-R-CNN \cite{He2017}, U-Net++ \cite{Zhou2018}; 
	and existing nuclei segmentation approaches, such as, HoVer-Net \cite{Graham2019}, 
	multi-organ nuclei segmentation, referred to as MoNS in this study \cite{Mahmood2020}, Stardist \cite{Weigert2022}, 
	WNSeg \cite{Liu2022}, Swin-MIL \cite{Qian2022} and BoNuS \cite{Lin2024}.
\end{itemize}

\par In this study, the comparative performance of SiliCoN model and other existing approaches in color normalization is analyzed using the UCSB breast cancer cell data set \cite{Gelasca2008}, published by the University of California, Santa Barbara. The UCSB breast cancer data contains a total of $58$ H\&E stained images: $26$ malignant cell and $32$ benign cell images. This data set is comprised of $10$ biopsy sets: $9$ of the sets contain $6$ images each and one set has $4$ images. Each UCSB data set image is stored in 24 bit non-linear RGB format and has a resolution of $896 \times 768$.

\par The effectiveness of the proposed SiliCoN model and several existing methods in nuclei segmentation is analyzed using H\&E stained TCGA image data set \cite{Kumar2017}. This tissue image set contains $1000\times 1000$ image patches extracted from $30$ whole slide images (WSIs) that were downloaded from The Cancer Genomic Atlas (TCGA). More than 21,000 labelled nuclei are present in this data set. 

\par For training the proposed SiliCoN model, corresponding to $15$ training images, $2,535$ overlapping $256\times 256$ size image patches are used, while $507$ overlapping patches are used for validation and $192$ non-overlapping image patches constitute the test set. 
For training the SiliCoN model, NVIDIA RTX A4000 ($16$ GB storage and $6144$ CUDA cores), is used .

\par As stain representative vectors extracted from the images within a particular biopsy set are expected to be in a close 
vicinity with each other, element-wise standard deviation is considered to analyze the 
performance of different variants of the SiliCoN model and other 
existing stain estimation approaches. On the other hand, for analyzing the effectiveness 
of the SiliCoN model over different existing methods in color normalization, normalized median intensity (NMI) \cite{Nyul2000}, between-image color constancy (BiCC) index and within-set color constancy (WsCC) index \cite{Maji2020} are utilized. In this study, to analyze the performance of 
different methods in segmenting nuclei regions, standard evaluation indices: Dice coefficient, Jaccard score, precision and 
recall, have been used.

\subsection{Performance in Stain Color Normalization}
The efficacy of the SiliCoN model over other existing approaches in stain vector estimation as well as color normalization is established through 
the following analyses:

\begin{figure}[h]
	\centerline{\includegraphics[width=3.6in,height=2.02in]{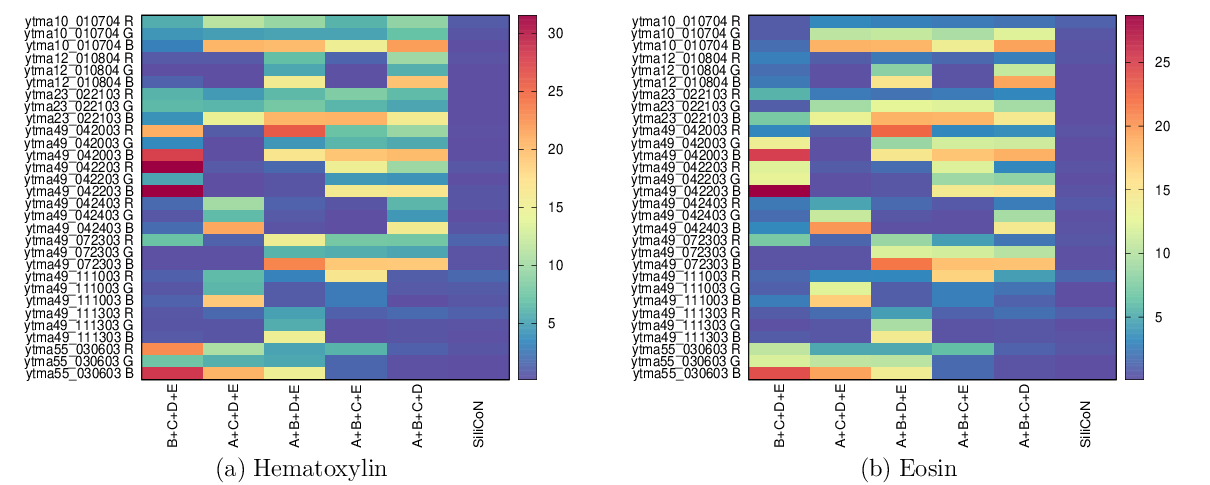}}
	\caption{Comparative performance analysis of SiliCoN 
		model and all combinations of five constituent terms of the objective function 
		in stain vector estimation. 
	}
	\label{SiliCoN_Ablation_heatmap}
\end{figure}

\begin{figure}[h]
	\centerline{\includegraphics[width=3.55in,height=1.42in]{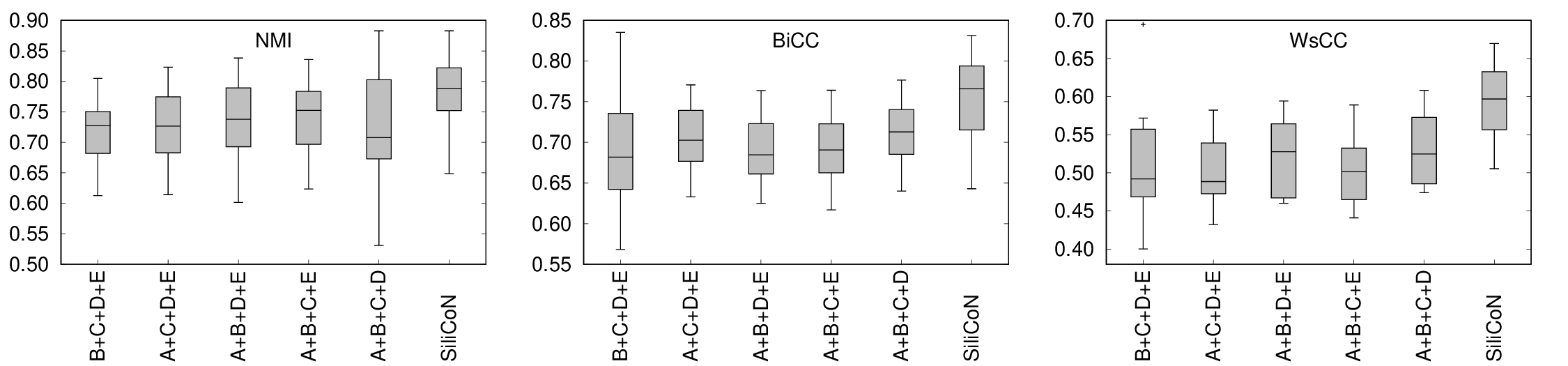}}
	\caption{Comparative performance analysis of SiliCoN 
		model and all combinations of five constituent terms of the objective function 
		in stain color normalization. 
	}
	\label{SiliCoN_Ablation_nmi_bicc_wscc}
\end{figure}

\subsubsection{Ablation Study}
\label{SiliCoN_ablationstudy}

\begin{table}[h]
	\begin{center}
		\caption{Statistical Significance Analysis with respect to Several variants and existing Methods}
		\label{SiliCoN_statistical_significance}
		\footnotesize{
			\begin{tabular}{|@{}c@{}|@{}c@{}|@{}c@{}|@{}c@{}|@{}c@{}|@{}c@{}|@{}c@{}|}\hline
				Different & \multicolumn{2}{c|}{{NMI}} & \multicolumn{2}{c|}{{BiCC}} & \multicolumn{2}{c|}{{WsCC}}\\\cline{2-7}
				Methods & ~Paired-\textit{t}~ & Wilcoxon & ~Paired-\textit{t}~ & Wilcoxon & ~Paired-\textit{t}~ & Wilcoxon \\\hline \hline
				B+C+D+E	&	4.98E-17	&	2.57E-11	&	4.44E-09	&	1.58E-07	&	9.19E-04	&	4.67E-03	\\\hline	
				A+C+D+E	&	2.40E-14	&	1.75E-11	&	7.55E-16	&	4.24E-11	&	4.34E-07	&	2.53E-03	\\\hline	
				A+B+D+E	&	5.63E-11	&	1.20E-09	&	3.52E-19	&	1.85E-11	&	7.52E-06	&	2.53E-03	\\\hline	
				A+B+C+E	&	2.28E-10	&	3.77E-11	&	2.31E-20	&	1.85E-11	&	5.85E-06	&	2.53E-03	\\\hline	
				A+B+C+D	&	2.91E-10	&	2.63E-08	&	6.78E-19	&	2.40E-11	&	1.01E-07	&	2.53E-03	\\\hline	\hline
				Correlated	&	4.07E-12	&	4.02E-11	&	9.88E-19	&	1.75E-11	&	1.47E-05	&	2.53E-03	\\\hline	\hline
				ColTrans	&	9.23E-23	&	1.75E-11	&	8.62E-30	&	1.75E-11	&	9.81E-10	&	2.53E-03	\\\hline	
				PF	&	3.57E-21	&	2.53E-11	&	1.50E-24	&	1.75E-11	&	4.87E-08	&	2.53E-03	\\\hline	
				EPF	&	7.73E-23	&	1.85E-11	&	8.49E-25	&	1.75E-11	&	1.92E-08	&	2.53E-03	\\\hline	
				SCD	&	2.34E-16	&	7.43E-11	&	2.32E-23	&	1.95E-11	&	4.39E-05	&	2.53E-03	\\\hline	
				HTN	&	2.81E-11	&	7.76E-09	&	1.23E-15	&	1.17E-10	&	2.38E-04	&	3.46E-03	\\\hline	
				SPCN	&	1.80E-23	&	1.75E-11	&	1.45E-26	&	1.75E-11	&	9.45E-08	&	2.53E-03	\\\hline	
				SN-GAN	&	7.60E-15	&	5.67E-10	&	7.89E-23	&	1.85E-11	&	2.03E-06	&	2.53E-03	\\\hline	
				StainGAN	&	5.60E-25	&	1.75E-11	&	1.98E-30	&	1.75E-11	&	6.90E-09	&	2.53E-03	\\\hline	
				AST	&	2.36E-19	&	1.85E-11	&	3.51E-25	&	1.75E-11	&	7.96E-07	&	2.53E-03	\\\hline	
				$\text{RFCC}_{\text{vM}}$	&	5.74E-10	&	3.19E-08	&	3.40E-11	&	8.88E-09	&	1.07E-03	&	4.67E-03	\\\hline	
				TredMiL	&	3.16E-06	&	2.69E-07	&	4.26E-11	&	4.30E-09	&	1.71E-04	&	2.53E-03	\\\hline			
			\end{tabular}}
	\end{center}
\end{table}

Combining (\ref{SiliCoN_adversarial_loss}), (\ref{SiliCoN_reconstruction_loss}) and (\ref{SiliCoN_reconstruction_full}), it can be stated that, apart from the adversarial loss $J_{\rm Adv}$, the objective function of the SiliCoN model also contains five constituent terms. Let, these five terms be represented by: 
$A = -E_{Q(z_{c}, y, z_{\omega})}[\log P_{\mathbb{G}}(x \mid z_{c}, y, z_{\omega})]$, 
$B = D_{KL}[Q(z_{c}) \mid\mid P_{\mathbb{Z}_{c}}(z_{c})]$, 
$C = D_{KL}[Q(y, z_{\omega}) \mid\mid P_{\mathbb{YZ_{\omega}}}(y, z_{\omega})]$, 
$D = -E_{Q(z_{c}, y, z_{\omega})} [\log Q(z_{c}, y, z_{\omega})]$ and 
$E = l_{SSIM}(\mathbb{G}, \mathbb{E}_{c}, \mathbb{F}_{\phi}, \mathbb{E}_{\omega})$. 
To establish the importance of every constituent term, an ablation study is used in this study, where each constituent term is removed from the objective function and the respective performance 
of the model is observed in that scenario. Analyzing the heatmap presented in Fig.~\ref{SiliCoN_Ablation_heatmap}(a), it becomes evident 
that the SiliCoN model with all the constituent terms present in the objective function outperforms 
every other combination in the estimation of H-stain representative vector. Similarly, the 
heatmap representation provided in Fig.~\ref{SiliCoN_Ablation_heatmap}(b) highlights the fact that 
the proposed SiliCoN model performs better than every other combination of the constituent terms in representative vector estimation corresponding to E-stain also.

\par To analyze the effectiveness of the proposed SiliCoN model, the performance of the model in stain color normalization is compared with that of the five combinations of the 
aforementioned constituent terms using indices NMI, BiCC 
and WsCC, and respective results are presented in Fig.~\ref{SiliCoN_Ablation_nmi_bicc_wscc}. 
The boxplot representation in Fig.~\ref{SiliCoN_Ablation_nmi_bicc_wscc} depicts the fact that 
the SiliCoN model achieves highest median values with respect to all the 
quantitative indices. Again, the statistical significance of the 
SiliCoN model is analyzed in terms of computed p-values using one-tailed tests: paired-\textit{t} and Wilcoxon signed-rank. It is evident from reported p-values in Table \ref{SiliCoN_statistical_significance} that the proposed SiliCoN 
model performs significantly better than all other combination of aforementioned constituent terms of the objective function, considering a confidence level of 95\%.


\subsubsection{Independence Between $z_{c}$, and $y$ and $z_{\omega}$}

\begin{figure}[h]
	\centerline{\includegraphics[width=3.6in,height=2.02in]{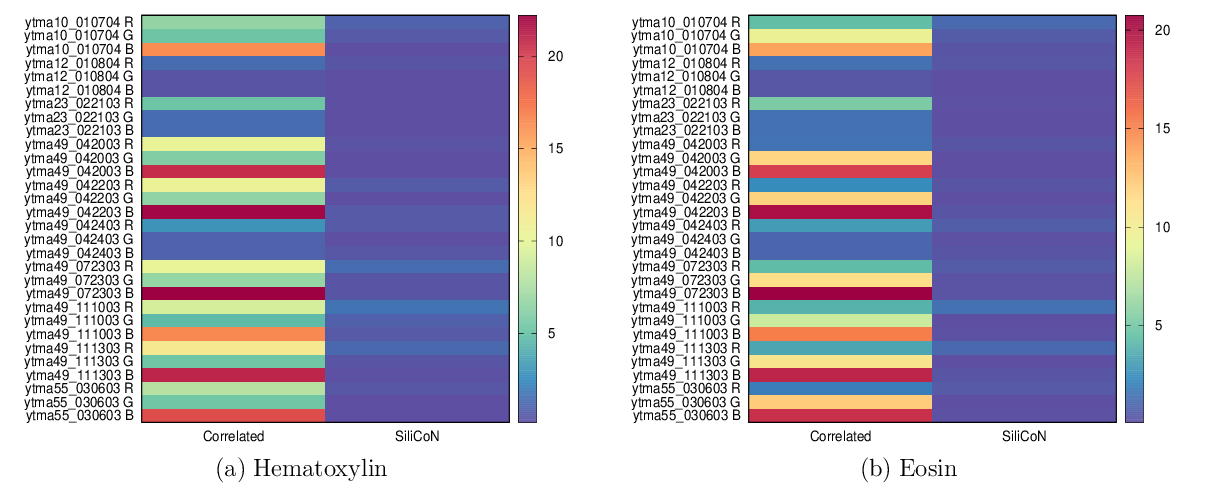}}
	\caption{Comparative performance analysis of SiliCoN 
		model and its counterpart, with correlated $z_{c}$, $y$ and $z_{\omega}$, in the estimation of stain representative vectors.}
	\label{SiliCoN_Correlated_heatmap}
\end{figure}

\begin{figure}[h]
	\centerline{\includegraphics[width=3.55in,height=1.22in]{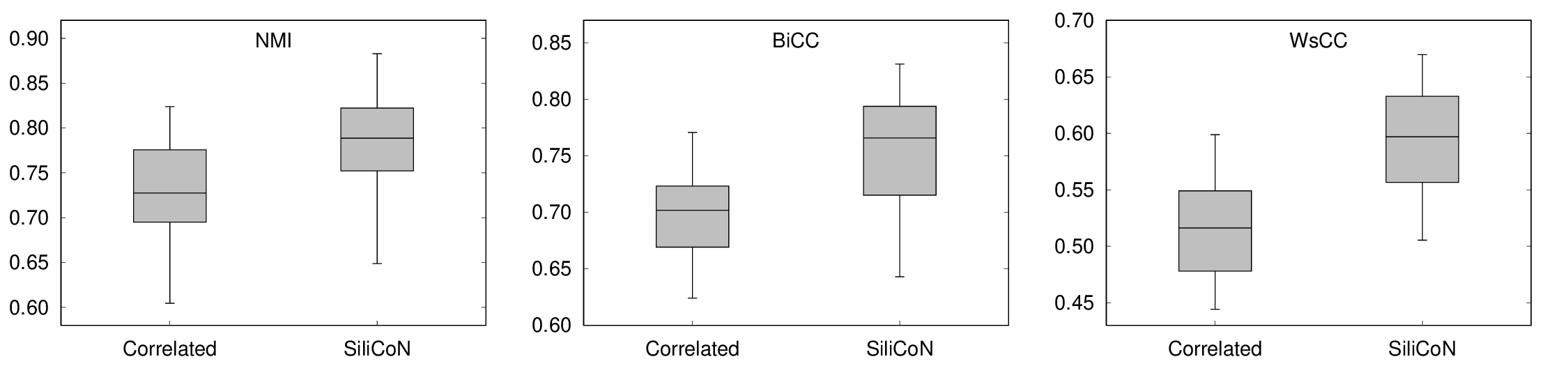}}
	\caption{Comparative performance analysis of SiliCoN 
		model and its counterpart, with correlated $z_{c}$, $y$ and $z_{\omega}$, in stain color normalization. 
	}
	\label{SiliCoN_Correlated_nmi_bicc_wscc}
\end{figure}

The proposed SiliCoN model is developed depending on the assumption that the latent color appearance code $z_{c}$ is independent of nuclei segmentation map $y$ as well as the embedding map $z_{\omega}$. 
To assess the effectiveness of the SiliCoN model in the estimation of stain representative vectors, the performance of SiliCoN model is compared with that of its counterpart where $z_{c}$, $y$ and $z_{\omega}$ are correlated, and 
the respective results are provided in 
Fig.~\ref{SiliCoN_Correlated_heatmap} through heatmap representation. From Fig.~\ref{SiliCoN_Correlated_heatmap}(a), it can be observed that in case H-stain, 
the SiliCoN model performs better than the aforementioned counterpart. Similarly, analyzing Fig.~\ref{SiliCoN_Correlated_heatmap}(b), it can be depicted that corresponding to E-stain vector estimation, the SiliCoN outperforms the counterpart with correlated $z_{c}$, $y$ and $z_{\omega}$.

\par To assess the effectiveness of SiliCoN in stain color normalization, 
the performance of the model is compared with that of the aforementioned counterpart. The boxplot representation provided in 
Fig.~\ref{SiliCoN_Correlated_nmi_bicc_wscc} depicts that the proposed SiliCoN model attains higher median values than that of the the counterpart with correlated $z_{c}$, $y$ and $z_{\omega}$ with respect to indices NMI, BiCC and WsCC. 
The SiliCoN model outperforms its counterpart with correlated $z_{c}$, $y$ and $z_{\omega}$ due to the fact that during mapping, the counterpart model loses significant 
amount of histological information as the modification or loss in $z_{c}$ affects both $y$ and $z_{\omega}$. The p-values presented in Table \ref{SiliCoN_statistical_significance} ensure that the SiliCoN model achieves statistically significant result than that of its 
counterpart with correlated $z_{c}$, $y$ and $z_{\omega}$.

\subsubsection{Comparison with Existing Approaches}

\begin{figure}[h]
	\centerline{\includegraphics[width=3.6in,height=2.25in]{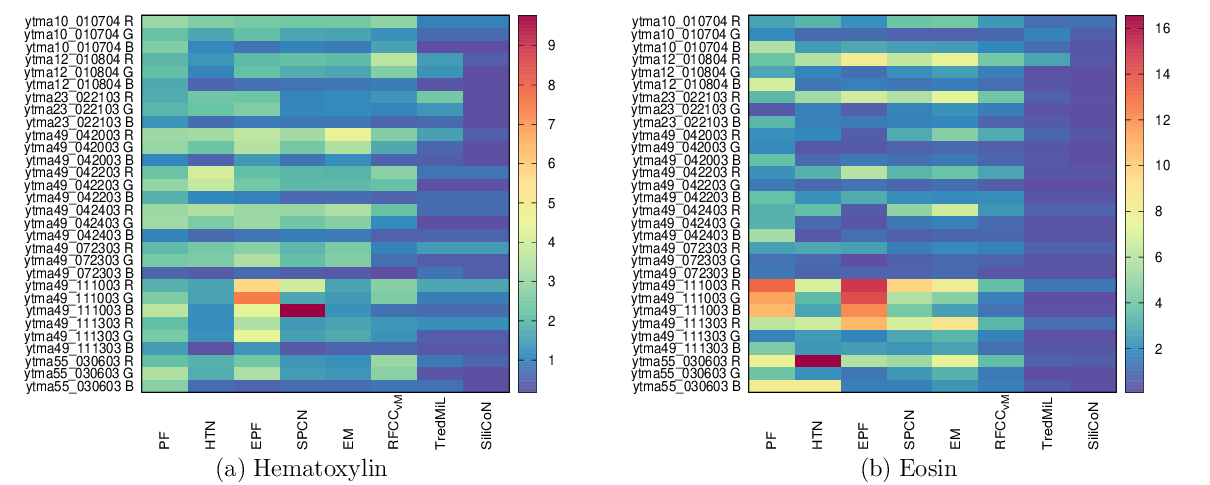}}
	\caption{Analysis of comparative performances of proposed SiliCoN model and several existing approaches: PF, HTN, EPF, SPCN, 
		EM, $\text{RFCC}_{\text{vM}}$ and TredMiL in the estimation of stain representative vectors.
	}
	\label{SiliCoN_Existing_heatmap}
\end{figure}

\begin{figure}[!h]
	\centerline{\includegraphics[width=3.6in,height=1.52in]{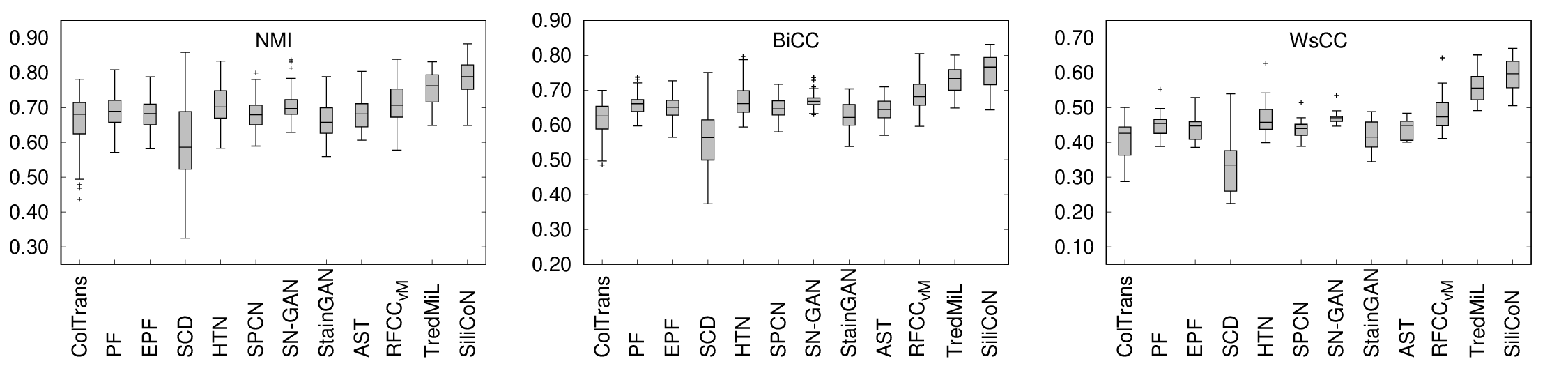}}
	\caption{Analysis of comparative performances of proposed SiliCoN model and 
		different existing approaches: ColTrans, PF, EPF, SCD, HTN, SPCN, 
		SN-GAN, StainGAN, AST , $\text{RFCC}_{\text{vM}}$ and TredMiL in the normalization of stain color appearance.
	}
	\label{SiliCoN_Existing_nmi_bicc_wscc}
\end{figure}

\begin{figure*}[!h]
	\centerline{\includegraphics[width=1.0\textwidth]{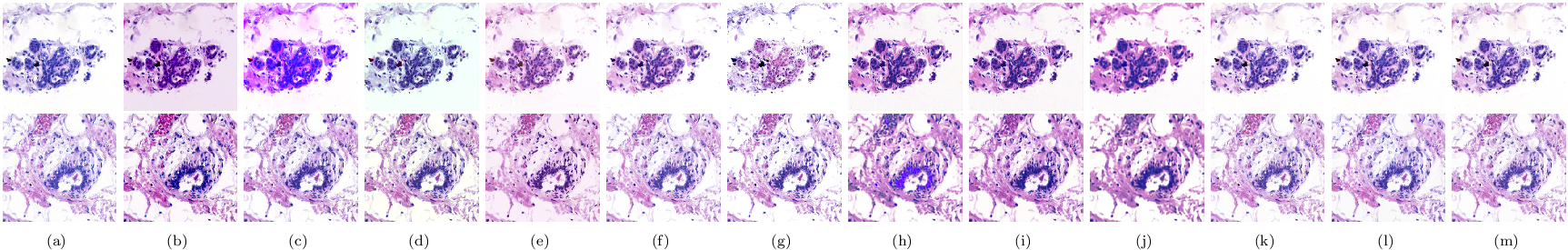}}
	\caption{(a) Original images of UCSB data; and stain color normalized images 
		obtained through several existing color normalization algorithms:
		(b) ColTrans, (c) PF, (d) EPF, (e) SCD, (f) HTN, (g) SPCN, (h) SN-GAN, 
		(i) StainGAN, (j) AST, (k) $\text{RFCC}_{\text{vM}}$, (l) TredMiL and (m) SiliCoN.}
	\label{SiliCoN_norm_qualitative}
\end{figure*}

Finally, to assess the efficacy of the SiliCoN model, its performance in the estimation of stain representative vectors is 
compared with that of different 
existing methods, and the respective results are provided in Fig.~\ref{SiliCoN_Existing_heatmap} via heatmap representations. 
From the heatmap presented in Fig.~\ref{SiliCoN_Existing_heatmap}(a), it can be noticed that the SiliCoN model outperforms the existing methods in case of H-stain 
representative vector estimation. It can also be seen in Fig.~\ref{SiliCoN_Existing_heatmap}(b) 
that SiliCoN attains lowest $\sigma$ values in 
most number of cases compared to the existing methods in representative vector estimation corresponding to the E-stain.

\begin{table}[!h]
	\begin{center}
		\caption{Comparative performance analysis in nuclei segmentation on TCGA Data: existing models vs SiliCoN}
		\label{SiliCoN_index_existing_kumar}
		\footnotesize{
			\begin{tabular}{|c|c|c|c|c|}\hline
				Methods	&	Dice	&	Jaccard	&	Precision	&	Recall	\\\hline	
				SiliCoN	&	\textbf{0.788345}	&	\textbf{0.653957}	&	\textbf{0.826117}	&	\textbf{0.786249}	\\\hline	\hline
				U-Net	&	0.645650	&	0.446128	&	0.624558	&	0.668216	\\\hline	
				Mask-R-CNN	&	0.747086	&	0.585659	&	0.799975	&	0.700757	\\\hline	
				U-Net++	&	0.773155	&	0.620294	&	0.796708	&	0.750955	\\\hline	
				HoVer-Net	&	0.744516	&	0.586780	&	0.811804	&	0.687529	\\\hline	
				MoNS	&	0.755509	&	0.614706	&	0.748012	&	0.763157	\\\hline	
				Stardist	&	0.743228	&	0.585713	&	0.811579	&	0.685495	\\\hline	
				WNSeg	&	0.773406	&	0.625322	&	0.813470	&	0.737104	\\\hline	
				Swin-MIL	&	0.749044	&	0.593266	&	0.801989	&	0.702656	\\\hline	
				BoNuS	&	0.784740	&	0.639496	&	0.806351	&	0.764257	\\\hline	
		\end{tabular}}
	\end{center}
\end{table}

\begin{figure*}[!h]
	\centerline{\includegraphics[width=0.98\textwidth]{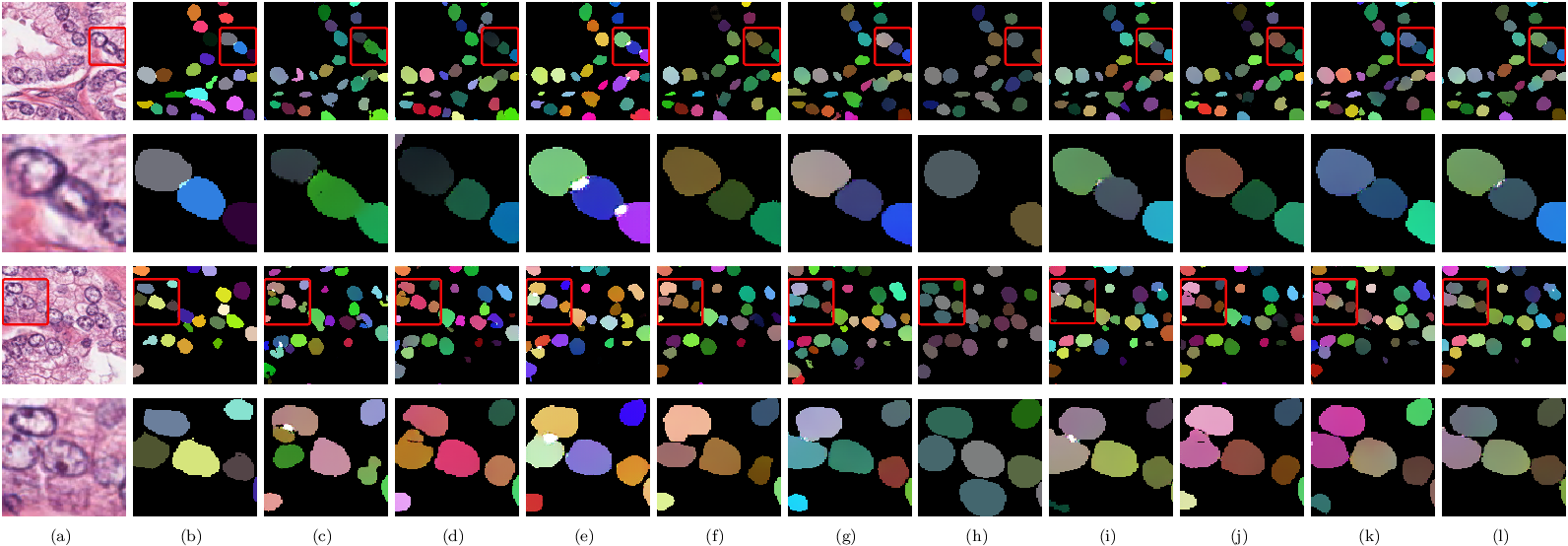}}
	\caption{(a) Original image patches of TCGA data, (b) ground-truth segmentation maps; and segmentation maps obtained through state-of-the-art segmentation approaches:
		(c) U-Net, (d) Mask-R-CNN, (e) U-Net++, (f) HoVer-Net, (g) MoNS, (h) Stardist, (i) Swin-MIL, (j) BoNuS and (k) SiliCoN. 
		Row 1 and row 3 present marked-up image patches, 
		and the zoomed-in regions corresponding to the marked-up 
		image patches in rows 1 and 3 are presented in rows 2 and 4, respectively.}
	\label{SiliCoN_existing_kumar}
\end{figure*}

\begin{table}[!h]
	\begin{center}
		\caption{Statistical significance analysis of different existing methods on TCGA data using 
			Paired-\textit{t} test }
		\label{SiliCoN_index_paired_t_kumar}
		\footnotesize{
			\begin{tabular}{|c|c|c|c|c|}\hline
				Methods	&	Dice 	&	Jaccard	&	Precision	&	Recall	\\\hline
				U-Net	&	3.89E-47	&	4.37E-49	&	1.14E-15	&	3.77E-51	\\\hline
				Mask-R-CNN	&	3.44E-31	&	1.01E-37	&	8.13E-81	&	1.83E-08	\\\hline
				U-Net++	&	8.79E-40	&	7.05E-41	&	4.70E-25	&	4.24E-26	\\\hline
				HoVer-Net	&	2.41E-52	&	5.58E-57	&	2.52E-12	&	3.10E-67	\\\hline
				MoNS	&	8.16E-28	&	2.12E-29	&	1.26E-60	&	1.93E-01	\\\hline
				Stardist	&	5.45E-34	&	1.63E-36	&	1.15E-04	&	7.20E-42	\\\hline
				WNSeg	&	1.16E-48	&	1.17E-49	&	4.20E-16	&	4.21E-63	\\\hline
				Swin-MIL	&	4.42E-51	&	2.95E-57	&	1.29E-16	&	5.34E-67	\\\hline
				BoNuS	&	3.84E-17	&	4.03E-18	&	9.68E-28	&	7.56E-26	\\\hline
		\end{tabular}}
	\end{center}
\end{table}

\begin{table}[!h]
	\begin{center}
		\caption{Statistical significance analysis of different existing methods on TCGA data using Wilcoxon signed-rank test}
		\label{SiliCoN_index_Wilcoxon_kumar}
		\footnotesize{
			\begin{tabular}{|c|c|c|c|c|}\hline
				Methods	&	Dice 	&	Jaccard	&	Precision	&	Recall	\\\hline
				U-Net	&	7.85E-32	&	8.61E-32	&	2.38E-14	&	6.25E-32	\\\hline
				Mask-R-CNN	&	1.30E-32	&	1.19E-32	&	1.86E-33	&	1.71E-02	\\\hline
				U-Net++	&	2.70E-29	&	2.44E-29	&	1.68E-25	&	1.42E-21	\\\hline
				HoVer-Net	&	3.22E-33	&	3.22E-33	&	5.00E-11	&	1.49E-33	\\\hline
				MoNS	&	7.62E-24	&	3.46E-24	&	1.47E-33	&	1.36E-01	\\\hline
				Stardist	&	7.12E-28	&	6.64E-28	&	1.14E-05	&	4.17E-30	\\\hline
				WNSeg	&	6.44E-32	&	6.34E-32	&	7.51E-15	&	1.62E-33	\\\hline
				Swin-MIL	&	1.92E-33	&	1.89E-33	&	1.84E-18	&	1.47E-33	\\\hline
				BoNuS	&	4.16E-16	&	1.98E-16	&	4.03E-24	&	7.74E-23	\\\hline
		\end{tabular}}
	\end{center}
\end{table} 

The boxplot representation provided in Fig.~\ref{SiliCoN_Existing_nmi_bicc_wscc} makes it evident 
that the SiliCoN model achieves better performance than the existing methods with respect to the quantitative indices used to evaluate the quality of color normalization: NMI, BiCC and WsCC. The reported p-values in Table \ref{SiliCoN_statistical_significance},
ensure the fact that the SiliCoN model is statistically more significant than the
existing color normalization methods. 
Fig.~\ref{SiliCoN_norm_qualitative} presents the qualitative performance analysis of 
several existing color normalization approaches. It can be concluded from the results presented in Table 
\ref{SiliCoN_statistical_significance}, Fig.~\ref{SiliCoN_Existing_nmi_bicc_wscc} and Fig.~\ref{SiliCoN_norm_qualitative} that the proposed SiliCoN model outperforms the state-of-the-art color 
normalization approaches in maintaining the color consistency after stain color normalization. 

\subsection{Performance in Nuclei Segmentation}

\par The comparative performance of the SiliCoN model in nuclei segmentation is analyzed against that of different existing approaches on TCGA data 
set and the corresponding results using standard segmentation evaluation indices, namely, Dice, Jaccard, precision and recall, are provided in Table 
\ref{SiliCoN_index_existing_kumar}. Assessing the values 
reported in Table \ref{SiliCoN_index_existing_kumar}, it becomes evident that, with respect 
to all the evaluation indices, the proposed SiliCoN 
model performs better than all the existing algorithms in nuclei segmentation. 
The qualitative performance of the SiliCoN model in nuclei segmentation, along with a comparison with the state-of-the-art approaches is presented in Fig. \ref{SiliCoN_existing_kumar}. 
The p-values reported in
Table \ref{SiliCoN_index_paired_t_kumar} depict the fact that SiliCoN performs significantly better in all the cases with respect paired-\textit{t} test. 
Again, analyzing the p-values presented in Table \ref{SiliCoN_index_Wilcoxon_kumar}, it is evident 
that SiliCoN performs statistically more significantly than all the existing methods in all the cases with respect Wilcoxon signed-rank test also. 

\section{Conclusion and Future Direction}
\label{SiliCoN_conclusion}
The problem of segmenting nuclei structures from histological images, in the presence of impermissible color variation within and between the histological images, is of utmost importance as the color variation among stained tissue images affect the performance of different nuclei segmentation approaches. In this context, the most impactful contribution of the paper is introducing a novel method, named SiliCoN, for simultaneously segmenting nuclei structures and normalizing color appearance of histological images. For addressing the stain overlap property of associated histochemical reagents, a mixture of truncated normal distributions is incorporated as the prior 
for latent color appearance code in the proposed SiliCoN model. 
Both the quantitative and qualitative results provided in the paper ensure the fact that 
SiliCoN outperforms existing approaches in nuclei segmentation as well as stain color normalization. The existing color normalization methods are also outperformed by the proposed SiliCoN model, as per as stain overlap handling, and ensuring within and between-image color consistency of nuclei regions after stain color normalization are concerned. The results reported in the paper also depicts the fact that nuclei segmentation on color normalized histological images enhances the 
segmentation accuracy. In future, more emphasis will be given to improve the nuclei segmentation model within the simultaneous framework in order to improve the performance of both tasks: nuclei segmentation and stain color normalization. 

\bibliographystyle{IEEEtran}
\bibliography{SiliCoN}

\begin{thebibliography}{10}
\providecommand{\url}[1]{#1}
\csname url@rmstyle\endcsname
\providecommand{\newblock}{\relax}
\providecommand{\bibinfo}[2]{#2}
\providecommand\BIBentrySTDinterwordspacing{\spaceskip=0pt\relax}
\providecommand\BIBentryALTinterwordstretchfactor{4}
\providecommand\BIBentryALTinterwordspacing{\spaceskip=\fontdimen2\font plus
\BIBentryALTinterwordstretchfactor\fontdimen3\font minus
  \fontdimen4\font\relax}
\providecommand\BIBforeignlanguage[2]{{%
\expandafter\ifx\csname l@#1\endcsname\relax
\typeout{** WARNING: IEEEtran.bst: No hyphenation pattern has been}%
\typeout{** loaded for the language `#1'. Using the pattern for}%
\typeout{** the default language instead.}%
\else
\language=\csname l@#1\endcsname
\fi
#2}}

\bibitem{Reinhard2001}
{E. Reinhard et~al.}, ``{Color Transfer Between Images},'' \emph{IEEE Computer
  Graphics and Applications}, vol.~21, no.~5, pp. 34--41, 2001.

\bibitem{Ruderman1998}
{D. L. Ruderman et~al.}, ``{Statistics of Cone Responses to Natural Images:
  Implications for Visual Coding},'' \emph{Journal of the Optical Society of
  America A}, vol.~15, no.~8, pp. 2036--2045, 1998.

\bibitem{Macenko2009}
{M. Macenko et~al.}, ``{A Method for Normalizing Histology Slides for
  Quantitative Analysis},'' in \emph{Proceedings of IEEE International
  Symposium on Biomedical Imaging: From Nano to Macro}, 2009, pp. 1107--1110.

\bibitem{Li2015}
{X. Li and K. N. Plataniotis}, ``{A Complete Color Normalization Approach to
  Histopathology Images Using Color Cues Computed from Saturation-Weighted
  Statistics},'' \emph{IEEE Transactions on Biomedical Engineering}, vol.~62,
  no.~7, pp. 1862--1873, 2015.

\bibitem{Vahadane2016}
{A. Vahadane et~al.}, ``{Structure-Preserving Color Normalization and Sparse
  Stain Separation for Histological Images},'' \emph{IEEE Transactions on
  Medical Imaging}, vol.~35, no.~8, pp. 1962--1971, 2016.

\bibitem{Bentaieb2018}
{A. Bentaieb and G. Hamarneh}, ``{Adversarial Stain Transfer for Histopathology
  Image Analysis},'' \emph{IEEE Transactions on Medical Imaging}, vol.~37,
  no.~3, pp. 792--802, 2018.

\bibitem{Zanjani2018}
{F. G. Zanjani et~al.}, ``{Stain Normalization of Histopathology Images Using
  Generative Adversarial Networks},'' in \emph{Proceedings of IEEE
  International Symposium on Biomedical Imaging (ISBI)}, 2018, pp. 573--577.

\bibitem{Shaban2019}
{M. T. Shaban et~al.}, ``{StainGAN: Stain Style Transfer for Digital
  Histological Images},'' in \emph{Proceedings of IEEE International Symposium
  on Biomedical Imaging (ISBI)}, 2019, pp. 953--956.

\bibitem{Zhu2017}
{J.-Y. Zhu et~al.}, ``{Unpaired Image-to-Image Translation using
  Cycle-Consistent Adversarial Networks},'' in \emph{Proceedings of IEEE
  International Conference on Computer Vision (ICCV)}, 2017, pp. 2242--2251.

\bibitem{Li2017}
{X. Li and K. Plataniotis}, ``{Circular Mixture Modeling of Color Distribution
  for Blind Stain Separation in Pathology Images},'' \emph{IEEE Journal of
  Biomedical and Health Informatics}, vol.~21, no.~1, pp. 150--161, 2017.

\bibitem{Maji2020}
{P. Maji and S. Mahapatra}, ``{Circular Clustering in Fuzzy Approximation
  Spaces for Color Normalization of Histological Images},'' \emph{IEEE
  Transactions on Medical Imaging}, vol.~39, no.~5, pp. 1735--1745, 2020.

\bibitem{Ronneberger2015}
O.~R. et~al., ``{U-Net: Convolutional Networks for Biomedical Image
  Segmentation},'' in \emph{Proceedings of Medical Image Computing and
  Computer-Assisted Intervention}, 2015, pp. 234--241.

\bibitem{Zhou2018}
{Z. Zhou et~al.}, ``{UNet++: A Nested U-Net Architecture for Medical Image
  Segmentation},'' in \emph{Proceedings of Deep Learning in Medical Image
  Analysis and Multimodal Learning for Clinical Decision Support}, 2018, pp.
  3--11.

\bibitem{He2017}
{K. He et~al.}, ``{Mask R-CNN},'' in \emph{Proceedings of IEEE International
  Conference on Computer Vision}, 2017, pp. 2961--2969.

\bibitem{Graham2019}
{S. Graham et~al.}, ``{Hover-Net: Simultaneous Segmentation and Classification
  of Nuclei in Multi-Tissue Histology Images},'' \emph{Medical Image Analysis},
  vol.~58, p. 101563, 2019.

\bibitem{Weigert2022}
{M. Weigert and U. Schmidt}, ``{Nuclei Instance Segmentation and Classification
  in Histopathology Images with Stardist},'' in \emph{Proceedings of IEEE
  International Symposium on Biomedical Imaging Challenges}, 2022, pp. 1--4.

\bibitem{Qian2022}
{Z. Qian et~al.}, ``{Transformer Based Multiple Instance Learning for Weakly
  Supervised Histopathology Image Segmentation},'' in \emph{Proceedings of
  Medical Image Computing and Computer Assisted Intervention}, 2022, pp.
  160--170.

\bibitem{Lin2024}
{Y. Lin et~al.}, ``{BoNuS: Boundary Mining for Nuclei Segmentation with Partial
  Point Labels},'' \emph{IEEE Transactions on Medical Imaging}, pp. 1--11,
  2024.

\bibitem{Mahapatra2023}
{S. Mahapatra and P. Maji}, ``{Truncated Normal Mixture Prior Based Deep Latent
  Model for Color Normalization of Histology Images},'' \emph{IEEE Transactions
  on Medical Imaging}, vol.~42, no.~6, pp. 1746--1757, 2023.

\bibitem{Prusty2023}
{M. R. Prusty et~al.}, ``{Nuclei Segmentation in Histopathology Images Using
  Structure-Preserving Color Normalization Based Ensemble Deep Learning
  Frameworks},'' \emph{Computers, Materials \& Continua}, vol.~77, no.~3, 2023.

\bibitem{Martos2023}
{O. Martos et~al.}, ``{Optimized Detection and Segmentation of Nuclei in
  Gastric Cancer Images Using Stain Normalization and Blurred Artifact
  Removal},'' \emph{Pathology - Research and Practice}, vol. 248, p. 154694,
  2023.

\bibitem{Mahbod2024}
{A. Mahbod et~al.}, ``{Improving Generalization Capability of Deep
  Learning-based Nuclei Instance Segmentation by Non-deterministic Train Time
  and Deterministic Test Time Stain Normalization},'' \emph{Computational and
  Structural Biotechnology Journal}, vol.~23, pp. 669--678, 2024.

\bibitem{Mahapatra2024}
{S. Mahapatra}, ``{Nuclei Segmentation and Color Normalization of Histological
  Images: Rough-Fuzzy Circular Clustering to Deep Generative Modeling},''
  Doctoral Dissertation, Indian Statistical Institute, December, 2024.

\bibitem{Jin2020}
{Q. Jin et~al.}, ``{RA-UNet: A Hybrid Deep Attention-Aware Network to Extract
  Liver and Tumor in CT Scans},'' \emph{Frontiers in Bioengineering and
  Biotechnology}, vol.~8, p. 605132, 2020.

\bibitem{Zhou2004}
{W. Zhou et~al.}, ``{Image Quality Assessment: From Error Visibility to
  Structural Similarity},'' \emph{IEEE Transactions on Image Processing},
  vol.~13, no.~4, pp. 600--612, 2004.

\bibitem{McCann2014}
{M. T. McCann et~al.}, ``{Algorithm and Benchmark Dataset for Stain Separation
  in Histology Images},'' in \emph{Proceedings of IEEE International Conference
  on Image Processing}, 2014, pp. 3953--3957.

\bibitem{Khan2014}
{A. M. Khan et~al.}, ``{A Nonlinear Mapping Approach to Stain Normalization in
  Digital Histopathology Images Using Image-Specific Color Deconvolution},''
  \emph{IEEE Transactions on Biomedical Engineering}, vol.~61, no.~6, pp.
  1729--1738, 2014.

\bibitem{Mahmood2020}
{F. Mahmood et~al.}, ``{Deep Adversarial Training for Multi-Organ Nuclei
  Segmentation in Histopathology Images},'' \emph{IEEE Transactions on Medical
  Imaging}, vol.~39, no.~11, pp. 3257--3267, 2020.

\bibitem{Liu2022}
W.~Liu, Q.~He, and X.~He, ``{Weakly Supervised Nuclei Segmentation Via Instance
  Learning},'' in \emph{Proceedings of IEEE International Symposium on
  Biomedical Imaging}, 2022, pp. 1--5.

\bibitem{Gelasca2008}
{E. D. Gelasca et~al.}, ``{Evaluation and Benchmark for Biological Image
  Segmentation},'' in \emph{Proceedings of IEEE International Conference on
  Image Processing}, 2008, pp. 1816--1819.

\bibitem{Kumar2017}
{N. Kumar et~al.}, ``{A Dataset and a Technique for Generalized Nuclear
  Segmentation for Computational Pathology},'' \emph{IEEE Transactions on
  Medical Imaging}, vol.~36, no.~7, pp. 1550--1560, 2017.

\bibitem{Nyul2000}
{L. G. Nyul et~al.}, ``{New Variants of A Method of MRI Scale
  Standardization},'' \emph{IEEE Transactions on Medical Imaging}, vol.~19,
  no.~2, pp. 143--150, 2000.

\end{thebibliography}
\end{document}